\RequirePackage{fix-cm}
\documentclass[twocolumn]{svjour3}          
\smartqed  
\usepackage{graphicx}
\usepackage{amsmath}
\usepackage{amsfonts}
\usepackage{amssymb}
\usepackage[titletoc,title]{appendix}
\usepackage{color}
\usepackage{epstopdf}
\usepackage{float}
\usepackage{caption}
\usepackage{subfig}
\usepackage{rotating}
\usepackage{enumerate}

\newcommand{\fref}[1]{Fig.~\ref{#1}}
\newcommand{\sref}[1]{Sec.~\ref{#1}}

\newcommand{\eref}[1]{Eq.~(\ref{#1})}

\begin{document}
\title{Unphysical features in the application of the Boltzmann collision operator in the time dependent modelling of quantum transport}

\author{Z. Zhan, E. Colom\'{e}s and  X. Oriols }
\institute{Z. Zhan, E. Colom\'{e}s and  X. Oriols \at
              Departament d\rq{}Enginyeria Electr\`{o}nica, Universitat Aut\`{o}noma de Barcelona, Spain. \\
              \email{xavier.oriols@uab.es}}

\date{\today}

\maketitle

\begin{abstract}

In this work, the use of the Boltzmann collision operator for dissipative quantum transport is analyzed. Its mathematical role on the description of the time-evolution of the density matrix during a collision can be understood as processes of adding and subtracting states. We show that unphysical results can be present in quantum simulations when the old states (that built the density matrix associated to an open system before the collision) are different from the additional states generated by the Boltzmann collision operator. As a consequence of the Fermi Golden rule, the new generated sates are usually eigenstates of the momentum or kinetic energy. Then, the different time-evolutions of old and new states involved in a collision process can originate negative values of the charge density, even longer after the collision. This unphysical feature disappears when the Boltzmann collision operator generates states that were already present in the density matrix of the quantum system before the collision. Following these ideas, in this paper, we introduce an algorithm that models phonon-electron interactions through the Boltzmann collision operator without negative values of the charge density. The model only requires the exact knowledge, at all times, of the states that build the density matrix of the open system.

\end{abstract}


\section{Introduction}
\label{sec1}

As it is well known, a quantum mechanical system is ruled by the Schr\"odinger equation when we deal with a closed system whose initial state is perfectly well known\footnote{When the initial state of a closed system is unknown and only the statistical probability of different initial states are known, the density operator, whose evolution is described by the Liouville equation, appears as a useful mathematical tool even for closed systems \cite{Dollfus2010}.}\cite{Cohen,OriolsBook}. However, this simple unitary Schr\"odinger equation cannot straightforwardly account for an open system interchanging energy and particles with the surroundings. From a practical perspective, an open system cannot be described by a pure state, but by a density matrix that describes mixed states (a statistical ensemble of several quantum states)\cite{Fischetti,Mizuta,Nedjalkov2011}. 

An electron device with an applied bias is indeed a quantum open system far from thermodynamic equilibrium. The electrons inside the active region of the device are part of the simulated degrees of freedom of the open system, while many other degrees of freedom (like the electrons in the battery, the phonons, etc.) are not included in the simulations. In the usual perturbation theory, the dynamics of the simulated electrons of interest are modeled through a well defined \emph{unperturbed} Hamiltonian. The interchange of energy between the simulated electrons and the environment\footnote{Although phonons and impurities are in the same physical space as the electrons, since their degrees of freedom are not simulated, we consider them ``outside'' of the system.}, like electron-phonon or electron-impurity interactions, is added into the  \emph{unperturbed} equation of motion as an additional term, the so-called \emph{collision} term \cite{Rossi2011}. The \emph{unperturbed} Hamiltonian term can be easily treated, while the \emph{collision} term requires typically  important approximations \cite{Rossi2011,Frensley1990,Nedjalkov2005}.

The consideration of the collision term in the modeling of quantum transport is of paramount importance because it converts a unitary and reversible equation of motion for the density matrix into a non-unitary and irreversible one, accounting for the phenomena of dissipation and decoherence. The proper treatment of such phenomena is mandatory for a realistic simulation of electron devices \cite{Jonasson,Querlioz2009,Oriols1996,Rossi2003}. There are many proposals for the collision operator in the literature \cite{Buttiker,Wang,Diosi,Querlioz2006,Frensley1986,Querlioz,Nedjalkov}. Some of such terms provide a quite accurate description of the phenomena at the price of a high requirement of computational resources. Because of that, such models are only applicable to very simple (idealized) systems. Other simpler versions of the collision term have a much wider practical applicability, but in some scenarios they can lead to unphysical results, in terms of negative values of the probability presence of electrons at some locations. 

In this work, we discuss the possibilities that the Boltzmann collision operator, applied to quantum transport, can produce unphysical results. We will discuss negative charge densities with examples using the density matrix and the  Wigner distribution function.

After this introduction, in \sref{sec2b}, we explain briefly the mathematical definitions of the density matrix and the Wigner distribution function, and some unphysical problems found in the literature related with different types of collision operators. Then, in section \sref{sec3}, we point out a problem inherent to the use of the Boltzmann collision operator in quantum systems and how to overcome it. In \sref{sec2} we show numerical results for a very simple system (an electron in a double barrier interacting with a phonon) where the Boltzmann collision operator leads to unphysical negative charge (probability presence) density. In \sref{sec4} we show numerically how this unphysical result can be removed by a careful treatment of the electron wave function before and after the scattering event. We conclude in \sref{sec5}.


\section{The density matrix and Wigner distribution function formalisms for dissipative transport}
\label{sec2b}

Hereafter, we explain briefly the equation of motion of the density matrix and the Wigner distribution function to treat dissipation in nanodevices. For the sake of simplicity, we will assume a mean field approximation that allows us to discuss quantum transport in terms of a set of non-interacting individual electrons. In particular, we will consider one-dimensional (1D) physical space $x$ or a 1D phase space $x-k$. There would be no fundamental difference in our conclusions, if a many-particle or three-dimensional (3D) physical space were considered.

\subsection{The density matrix formalism}

The \emph{natural} approach to model electrons in a open systems is the density operator, which can be written as a weighted sum of states\footnote{Each state $|\psi_j(t)\rangle$ can be interpreted as the single-particle wave function of an electron that evolves inside the open system independently (without quantum interfering with the other electrons). By construction, the ensemble value of an arbitrary operator $\hat A$ acting on an ensemble of states weighted by $p_j(t)$ provides the same ensemble values obtained from \eref{density} by $\langle A \rangle=tr({\hat A \hat \rho})$. }:
\begin{equation}
 \hat{\rho}(t)=\sum_j p_j(t) |\psi_j(t)\rangle \langle  \psi_j(t)|
 \label{density}
 \end{equation}
where $p_j(t)$  specifies the probability that the open system is described by the pure state $|\psi_j(t)\rangle$. The density operator expressed in the position representation is the following, 
 \begin{equation}
 \rho(x,x',t)= \langle x'|\hat{\rho}(t)|x\rangle =\sum_j p_j(t)\psi_j(x,t)\psi_j^\ast(x',t).
 \label{matrix}
 \end{equation}
For a Hamiltonian $\hat H=\hat H_0+\hat H'$, where $\hat H_0$ is the \emph{unperturbed} Hamiltonian and $\hat H'$ accounts for the collision of the electron with other particles \cite{Rossi2011,Rossi1998}, the equation of motion of the density operator $ \hat{\rho}$ is the so-called Liouville Von-Neumann equation:
\begin{align}
\frac{\partial \hat{\rho}}{\partial t}=\frac{1}{i\hbar}[\hat{H}, \hat{\rho}]=\frac{1}{i\hbar}[\hat{H_0}, \hat{\rho}]+\hat C[\hat{\rho}]
\label{liouville}
\end{align}
where the first part of the right hand side of \eref{liouville} is the \emph{unperturbed} evolution of the quantum system and the second part, the collision term $\hat{C}[\hat{\rho}]$, describes the effect of the interaction with the environment. For the effective mass approximation $m^\ast$, \eref{liouville} in the position representation becomes:
\begin{align}
i\hbar\frac{\partial \rho(x, x', t)}{\partial t}&=-\frac{\hbar^2}{2 m^\ast}\bigg( \frac{\partial^2}{\partial x^2} -\frac{\partial^2}{\partial x'^2}\bigg) \rho(x, x', t)  \nonumber \\
&+ \big( V(x)-V(x') \big) \rho(x, x', t) + C [\rho(x, x', t) ]
\label{liouville1}
\end{align}
where $V(x)$ is the potential energy term and $C$ is the matrix representation of $\hat C$ operator. 

An important quantity in the modeling of electrical devices can be obtained by the trace of the density matrix:
\begin{eqnarray}
 Q(x, t) =tr({\hat \rho(t)})=\sum_j p_j(t) |\psi_j(x,t)|^2.
\label{marg1}
\end{eqnarray}
The function $Q(x,t)$ is often referred in the literature as the charge (or probability presence) density of the system. By definition, either interpreted as a probability presence or as the system charge, negative values\footnote{Although not relevant for our work, strictly speaking, as seen for a pure state $Q(x,t)=|\psi(x,t)|^2$, \eref{marg1} represents the probability presence density. To avoid any misunderstanding, if one wants to interpret \eref{marg1} as charge, then we can also argue that we cannot accept a positive value of $-qQ(x,t)$ (with $-q=-1.6e^{-19}$ C the electron elementary charge) as the charge of electrons in an open quantum system} cannot be accepted in \eref{marg1}. 

\subsection{The Wigner distribution function formalism}

The Wigner distribution function is defined as the following Wigner-Weyl transform \cite{Wigner1932}  of $\rho(x,x',t)$ in \eref{matrix}:

\begin{eqnarray}
F_W(x, k, t) = \frac{1}{2\pi \hbar} \int  \rho(x+\frac{x'}{2},x-\frac{x'}{2},t) e^{-i k x'} \mathrm{d}x'.
\label{wigner1}
\end{eqnarray}

The time evolution of the Wigner distribution function can be directly derived from \eref{liouville}. Therefore, the transport equation for the Wigner distribution function in \eref{wigner1} can be written as a sum of a term given by the operator $\hat{L}_W\left[ F_W(x, k, t)\right]$ plus a generic collision term $\hat{C}_W \left[ F_W(x, k, t)\right]$ as: 
\begin{equation}
\label{Wigner_transport}
\frac{\partial F_W(x, k, t)}{\partial t}= \hat{L}_W\left[ F_W(x, k, t)\right] +\hat{C}_W\left[ F_W(x, k, t)\right]. 
\end{equation}
The term $\hat{L}_W\left[ F_W(x, k, t)\right]$ can be written as:
\begin{align}
\label{Liouville_superoperator}
\hat{L}_W\left[ F_W(x, k, t)\right]&=-\frac{\hbar k}{m^\ast}\frac{\partial F_W(x, k, t) }{\partial x}\nonumber\\
&-\frac{2}{\pi \hbar}\int_{-\infty}^{\infty}\mathrm{d}k'\int_0^\infty \mathrm{d}x' e^{[-i(k-k')2x']} \nonumber\\
&\qquad \times [V(x+x')-V(x-x')]F_W(x, k, t)
\end{align}
under the effective mass approximation. The other collision term $\hat{C}_W \left[ F_W(x, k, t)\right]$ has many different practical implementations (based on different approximation).

The charge density in \eref{marg1} can also be obtained by integrating the Wigner distribution over all momenta:
\begin{eqnarray}
 Q(x, t) =\hbar \int F_{W}(x, k, t) dk.
\label{marg2}
\end{eqnarray}

\subsection{Unphysical negative charge density found in the literature for quantum transport models}

\begin{table*}[ht]
\centering
\begin{tabular}{|p{0.3\linewidth}|p{0.52\linewidth}|p{0.1\linewidth}|}
\hline
\textbf{Collision term} & \textbf{Unphysical problem} & \textbf{Reference} \\
\hline
Non-Markovian treatment of collision & Without adding the Markovian approximation negative charge appears. & \cite{Rossi2016} \\
\hline
Barker-Ferry equation and Levinson equation  & The simulation time has to be small enough to avoid negative probability in the momentum space. & \cite{ferry2006}\\
\hline
Boltzmann collision operator, plus Fermi Golden rule  & It can give negative charge in some scenarios due to the fact that the``rates derived based on the Fermi Golden rule rely on a well-defined kinetic energy pre- and post-scattering states". & \cite{Jonasson}\\
\hline
Relaxation time approximation  & The final (thermodynamical) equilibrium state used in the approximation can be unknown or unphysical. & \cite{Oriols1996}\\
\hline
Any type of collision operator   & A phenomenological injection model (without inclusion of a collision term) can provide negative charge. & \cite{Rossi2003}\\
\hline
\end{tabular}
\centering
\caption{Some unphysical behaviors found in the literature when modeling dissipative quantum transport with the density matrix or Wigner distribution function formalisms using different implementations of the collision operator.}
\label{table0}
\end{table*}

In the literatures \cite{DAmato,Shifren,Jonasson2014,Polkovnikov,Smithey,Nedjalkov2013}, there is a large list of different approaches that are used to define the collision operator, either in the density matrix formalism $\hat{C}[\hat{\rho}]$ or in the Wigner formalism $\hat {C}_W \left[ F_W \right]$. As we said, the proper treatment of the interaction of electrons with the environment is mandatory for a realistic simulation of electron devices. However, since a direct simulation of all degrees of freedom is computationally inaccessible, the collision term cannot be treated exactly and some approximations are required. 
Such approximations do not only imply deviations from the exact result but, in some circumstances, they imply that the simulated results are unphysical (for example, the negative values of the charge density discussed in this paper).

In Table \ref{table0} we list some recent works of the literature explaining the negative values of the charge density, obtained from quantum transport formalism dealing with the density matrix or the Wigner distribution function, with different implementations of the collision terms. Thus, there are many different reasons that explain why we can obtain negative charge densities. One can realize the important conceptual and practical difficulties of the proper modeling of the collision term with the recent work of Rossi and co-workers \cite{Rossi2016}. There, the authors show that a quite detailed treatment (including only a mean-field approximation) of electron-phonon scattering in the density matrix formalism leads to negative values of the charge. Surprisingly, such unphysical results disappear when a simpler  treatment (including mean-field and Markovian approximations) is considered   \cite{Rossi2016}. Similarly, a detailed treatment of scattering within the Barker-Ferry equation (or the Levinson equation) with a generalized Wigner distribution function formalism shows the relevance of the simulation times in the description of the collision interaction. For such approaches, unphysical results appear if the simulation time is not short enough (such limitation being related to the first order perturbation done in the development of the electron-phonon coupling) \cite{ferry2006}. The implementation of the spatial boundaries conditions on the equation of motion of the density matrix at the borders of the simulation box can also be the origin of some unphysical results \cite{Rossi2003}. 

In this work, we analyze, indeed, a much simpler and widely used treatment of the collision term. We discuss why in some circumstances the use of the Boltzmann collision operator can produce unphysical results. It is important to notice here that the Boltzmann collision operator is usually implemented through the use of the Fermi Golden rule that determines the scattering rates\footnote{The Fermi Golden rule itself is developed under some approximations. First, the interaction time has to be small enough to make the first order perturbation development correct \cite{Cohen}. Second, the interaction time has to be large enough to ensure that the sinc function approaches a delta function in the development of the Fermi Golden rule \cite{Cohen}.} \cite{Dollfus2010,Jonasson2014,Barraud}. In our work, we do not discuss the range of validity of the approximation of the Fermi Golden rule, but only some inherent difficulties that can be found in the practical implementation of the Boltzmann collision operator in quantum simulators. To the best of our knowledge, although the idea was briefly mentioned in Ref. \cite{Jonasson}, the origin of the unphysical results presented in this paper has not been discussed in the literature explicitly.


\section{The problem and the solution}
\label{sec3}

The Boltzmann collision operator was initially proposed for classical systems \cite{bol}. For such systems, it has a very easy and understandable interpretation. The Boltzmann collision operator is just a rule for counting the number of electrons in and out of a volume of the phase space $\Delta V$ due to a collision. The total number of electrons at time $t+\Delta t$ in $\Delta V$ is equal to the previous number of electrons that were there at $t$, before the collision, plus the number of electrons that arrive at $\Delta V$ from outside due to collision, minus the number of electrons that leave $\Delta V$ during the collision.

Let us imagine a classical electron at $x_0$ with a velocity $v_0$ that interacts with another particle (for example, a phonon). Because of the interaction, the electron losses kinetic energy and its final velocity is $v_f$. For simplicity, we assume that the initial position remains unchanged. Such collision process can be easily modeled in terms of the previous Boltzmann collision operator. The initial classical distribution function in phase space before the collision (apart from constant factors), at time $t_0$, is:
\begin{eqnarray}
F_c(x,k,t_0)&=& \delta(x-x_0)\delta(k-k_0)
\end{eqnarray}
being $k_0$ the wave vector associated to $v_0$.  The Boltzmann collision operator generates the effect of the collision by subtracting an electron with momentum $k_0$ and adding a new electron with momentum $k_f$. The final classical distribution function in phase space after the collision, at time $t=t_0+\Delta t$, is: 
\begin{eqnarray}
F_c(x,k,t_0+\Delta t)&=&\delta(x-x_0)\delta(k-k_0)\nonumber\\
&-&\delta(x-x_0)\delta(k-k_0)\nonumber\\
&+&\delta(x-x_0)\delta(k-k_f)\nonumber\\&=&\delta(x-x_0)\delta(k-k_f). 
\end{eqnarray}
The final results is obviously $F_c(x,k,t_0+\Delta t)=\delta(x-x_0)\delta(k-k_f)$. Up to here, the discussion seems very trivial. However, let us emphasize that it has been relevant that the negative part of the distribution function generated by the Boltzmann collision operator $-\delta(x-x_0)\delta(k-k_0)$ is exactly compensated by the original positive one $\delta(x-x_0)\delta(k-k_0)$. In this sense, the use of the Boltzmann collision operator in classical systems will always be unproblematic. 

However, the application of the quantum version of the Boltzmann collision operator can be problematic because we have to add/subtract quantum states or wave functions, not point particles. As we will see next, the problem appears when we do not know the states that built the density matrix of the open system.  

We consider an electron device as an open system with $M$ electrons which are distributed in $N$ different states. Say, there are $M_1$ electrons described by state $|\psi_1 \rangle$ where we define $p_1=M_1/M$. There are $M_2$ electrons with probability $p_2=M_2/M$ described by the state $|\psi_2 \rangle$ and so on. We construct a mixed state through the density matrix that describe our open system according to \eref{density}:
\begin{eqnarray}
\hat{\rho}_B(t_0)= \sum_{i=1}^N p_i(t_0)|\psi_i(t_0) \rangle \langle\psi_i(t_0)|
 \label{wig1}
\end{eqnarray}
with the conditions $\sum_{i=1}^N p_i(t_0)=1$ and $\sum_{i=1}^N M_i=M$. Because of the interaction of one electron with a phonon, the Boltzmann collision operator will add a new final state of the electron $|\psi_{F}\rangle$ and will subtract another state associated to the electron $|\psi_O \rangle$. Then, the new density matrix after the scattering, at time $t_S=t_0+\Delta t$, will be: 
\begin{eqnarray}
\hat{\rho}(t_S)=\hat{\rho}_B(t_0)
-\frac{1}{M}|\psi_O \rangle \langle\psi_O|+\frac{1}{M} |\psi_{F} \rangle \langle\psi_{F}|
 \label{wig2}
\end{eqnarray}
In next \sref{PW}, we will show explicitly with the Wigner formalism how the effect of collisions modelled by the Boltzmann collision operator can effectively be written as  \eref{wig2}. The problem with the expression (\ref{wig2}), due to Boltzmann collision operator, is that if we subtract an state $|\psi_O \rangle=|\psi'_2 \rangle$ that is not present in the density matrix before the collision, $\hat{\rho_B}(t_0)$, then we cannot simplify the density matrix to remove the negative sign that appears in the second term of the right hand side of \eref{wig2}. Then, by a simple computation, the new expression of the charge density with \eref{marg1} using the density operator in \eref{wig2} is: 
\begin{eqnarray}
Q(x,t)&=&\sum_{i=1}^N p_i|\psi_i(x,t)|^2\nonumber\\&-&\ \frac {1}{M}|\psi'_2(x,t)|^2+\frac {1}{M}|\psi_{F}(x,t)|^2
\label{q2}
\end{eqnarray}
This charge density is a sum of positive and negative terms. The dramatic problem with \eref{q2} is that, when the time-evolution of the negative term $\psi'_2(x,t)$ is not perfectly  balanced by the positive term $\psi_2(x,t)$ (or by other states that build $\hat \rho_B$) at every time and position, the possibility of getting negative values $Q(x,t)$ is opened.  

The solution of the unphysical result originated by a negative charge density is, in principle, quite simple. If we subtract a state $|\psi_O \rangle$ which is present in the density matrix $\hat{\rho_B}(t_0)$, for example, $|\psi_O \rangle=|\psi_2 \rangle $, Then, we can write the density matrix in \eref{wig2} at any time $t$ after the scattering time $t_S$ as:
\begin{eqnarray}
\hat{\rho}(t)&=& \sum_{i=1;i \neq 2}^N p_i(t)|\psi_i(t) \rangle \langle\psi_i(t)| \nonumber\\
&+&\frac{M_2-1}{M}|\psi_2(t) \rangle \langle\psi_2(t)|+\frac{1}{M}|\psi_{F}(t) \rangle \langle\psi_{F}(t)|
 \label{wig3}
\end{eqnarray}
The relevant point now is that, by construction, the term $(M_2-1)/M$ will be positive at any time $t$. Obviously, in the selection of the scattering process we have to ensure that $M_2 \geq1$, because if not, we would be subtracting a non existent state. If the condition $|\psi_O \rangle=|\psi_2 \rangle$ is satisfied during the collision, then, independently of the time-evolution of all the states, the charge density computed from \eref{marg1} using the density operator in \eref{wig3} is just a sum of positive terms:
\begin{eqnarray}
Q(x,t)&=&\sum_{i=1;i \neq 2}^N p_i|\psi_i(x,t)|^2\nonumber\\&+&\frac{M_2-1}{M}|\psi_2(x,t)|^2+\frac{1}{M}|\psi_{F}(x,t)|^2
\label{q1}
\end{eqnarray}
Let us mention, however, that this procedure requires a knowledge of the pure states that build the density matrix (or the Wigner distribution function) of our open system. This information is usually not available in most quantum transport simulations. An exception being the BITLLES simulator, where each electron inside the device has its own (conditional) wave function \cite{BITLLES}. 

\section{Numerical example of the problem}
\label{sec2}

In this section, we will show with numerical results the potential drawbacks of the combination of the Wigner distribution function and the Boltzmann collision operator discussed in the previous section. In some scenarios, such combination can lead to negative values of the charge density. Later, in \sref{sec4}, we will develop a novel scattering model (taking care of adding and subtracting states present in the density matrix of the open system) where the previous unphysical feature disappears by construction. 

\subsection{Boltzmann collision operator for Hamiltonian eigenstates}

The Boltzmann collision operator in the Wigner formalism is given in the literature \cite{Dollfus2010} by:
\begin{align}
\label{Boltzmann_operator}
&\hat{C}_W\left[ F_W(x, k, t)\right] = \nonumber\\
&\qquad\quad \frac{1}{2 \pi}\int_{-\infty}^{\infty}\{W_{kk'}F_W(x, k', t)-W_{k'k}F_W(x, k, t)\}\mathrm{d}k'
\end{align}  
where $W_{kk'}$ is the rate of scattering from the (\emph{unperturbed} Hamiltonian) eigenstate with  eigenvector $k'$ to the (\emph{unperturbed} Hamiltonian) eigenstate with eigenvector $k$. The transition probabilities $W_{kk'}$ are obtained from the Fermi Golden rule according to \cite{Nag}:
\begin{equation}
\label{scattering_rate_3D}
W_{\vec{k}\vec{k'}}=\frac{2\pi}{\hbar}|M_{\vec{k}\vec{k'}}|^2 \delta (E_{\vec{k}}- E_{\vec{k'}}\mp \hbar \omega)
\end{equation}
where $M_{\vec{k}\vec{k'}}$ are the matrix elements for the transitions from state $\vec{k'}$ to $\vec{k}$,  and $\omega$ is the frequency of the phonon for inelastic scattering. The bold symbols represents vectors in the 3D space. For technical reasons, since only one dimension is considered in our work, the 3D scattering rates must be ``projected '' onto the one-dimensional model to find $W_{kk'}$ defined in \eref{Boltzmann_operator}:
\begin{equation}
\label{scattering_rate}
W_{kk'}=\frac{\lambda_{T}^2}{(2\pi)^3}\int_{-\infty}^{\infty}\mathrm{d}^2k_\perp'\int_{-\infty}^{\infty}\mathrm{d}^2k_\perp W_{\vec{k}\vec{k'}}exp(-\frac{\lambda_{T}^2k_\perp^2}{2})
\end{equation}
where $k$ and $k'$ are now the 1D initial and final states respectively.\footnote{Here we assume that the distribution of electrons is Maxwellian with respect to the transverse wave vector $k_\perp$ of the initial state, and $\lambda_T$ is the spatial dimension factor given by:
\begin{equation}
\label{lambda_T}
\lambda_T=\frac{\hbar^2}{KTm^\ast}
\end{equation}
where $K$ is the Boltzmann constant and $T$ is the absolute temperature.} A very relevant point in our discussion is that the Fermi Golden rule \eref{scattering_rate_3D} forces us to use Hamiltonian eigenstates (of the Hilbert space without the interacting potential) to compute the matrix elements $M_{\vec{k}\vec{k'}}$.

In following \sref{PW}, we will show a simple example of the application of \eref{Wigner_transport} for a simple initial state and scattering rates $W_{kk'}$,  that, surprisingly, gives unphysical results in the form of negative charge density evaluated according to \eref{marg2}.


\subsection{Electron in a double barrier with a collision event}
\label{PW}

We note here that we are not focused on the simulation of realistic nanodevices, but only in showing with a very simple example an unexpected result when combining the Boltzmann collision operator and the Fermi Golden rule. The violation of the requirement $Q(x,t) \geq 0$ in only one simple system is enough to warn that such implementation of the collision operator can lead to unphysical results in more complex or realistic simulations. 

\begin{figure}[t!!!!!!]
\centering
\includegraphics[scale=0.5]{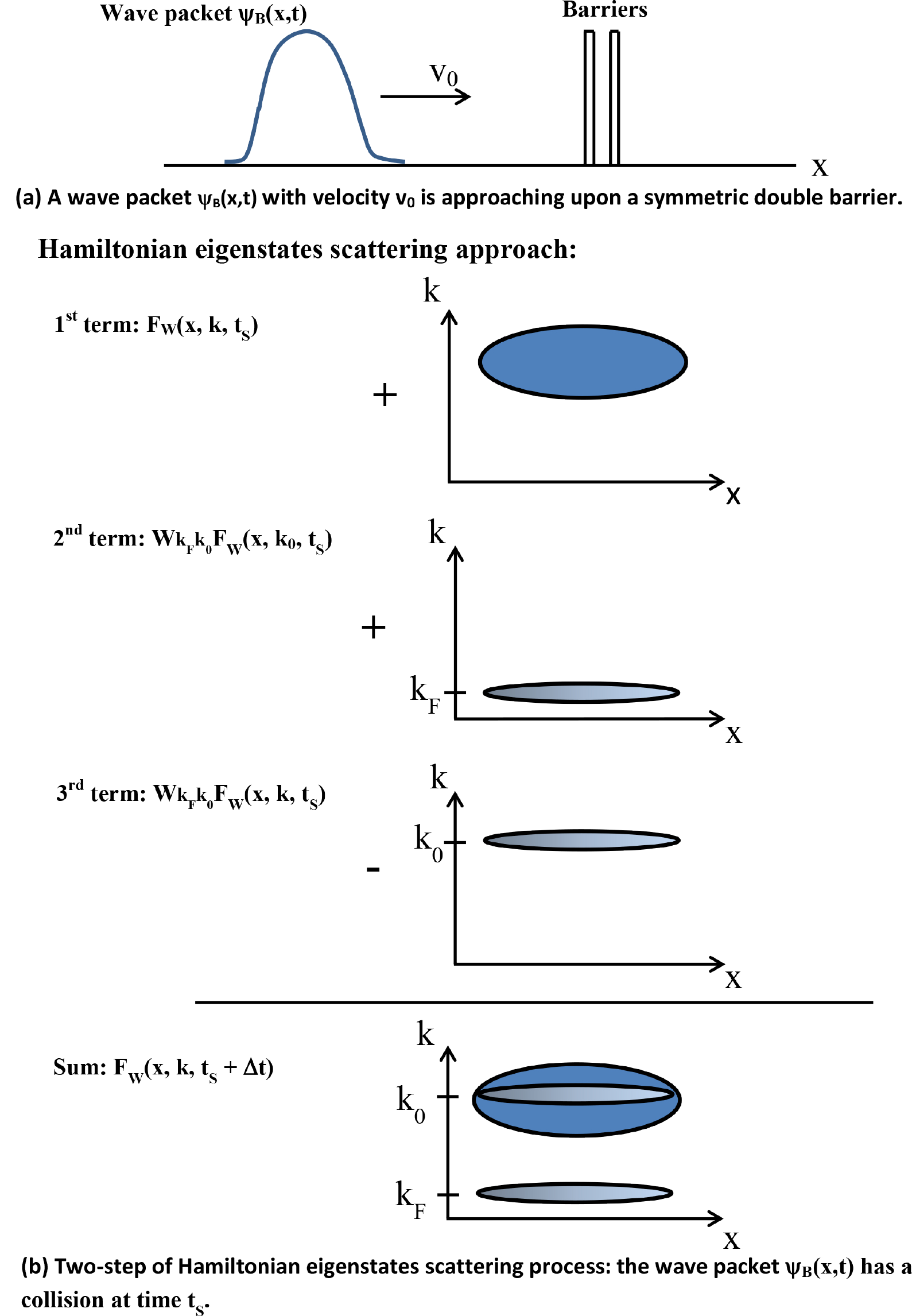}
\caption{Schematic representation of the two-step Hamiltonian eigenstates scattering process. (a) Simulation of a wave packet impinging on double barriers, the collision is performed at time $t_S$ before the wave packet touches the barriers. (b) A simple physical picture of the two-step  Hamiltonian eigenstates scattering process. }
\label{scattering_process1}
\end{figure}

As seen in the inset of Fig. \ref{scattering_process1}, we consider an electron that suffers a collision with a phonon while traveling through a typical double barrier potential. We consider a 1D Hilbert space with the following uniform grid $x_j=j\Delta x,\; \textrm{for}\; j=1,2,\dots M$ with $\Delta x=0.2$ nm the spatial step and $M=3000$ the number of grid points. The simulation box is large enough (it extends from $0$ till $600$ nm) to avoid any spurious interaction of the wave packet  with the spatial boundaries. In the simulation, the temporal step is $\Delta t=3$ fs. 

At the initial time $t_0$ we consider an arbitrary initial pure state $\langle x|\psi_B\rangle$ whose support fits perfectly inside the simulation box. Since we are interested in describing such system with the Wigner distribution function, the density matrix of this initial pure state is given by $\hat {\rho_B} =|\psi_B \rangle \langle \psi_B|$, and the Wigner distribution function just needs the Wigner-Weyl transform given by \eref{wigner1}. The time-evolution of the Wigner distribution function can be computed directly by solving the Schr\"{o}dinger equation (plus a Wigner-Weyl transform) or by solving the Wigner transport equation \eref{Wigner_transport} without the collision operator.  Then, at time $t_S$, a scattering process takes place according to the Boltzmann collision operator.  

It is usually assumed in the literature that the scattering process is sufficiently instantaneous\footnote{As indicated in \cite{Shankar}, such assumption is not always valid. In any case, the consideration of a larger time will not significantly change the drawbacks of the Boltzmann collision operator mentioned here.} that we can assume that the evolution of the Wigner distribution function from the $t_S$ till $t_S+\Delta t$ is:
\begin{align}
\label{discrete}
&\left. \frac{\partial F_W(x, k, t)}{\partial t}\right|_{t=t_S}\simeq\frac{F_W(x,k, t_S+\Delta t)-F_W(x,k,t_S)}{\Delta t}    \nonumber\\
&=\frac{1}{2\pi}\int \{ W_{kk'}F_W(x,k',t_S)-W_{k'k}F_W(x,k,t_S) \}\mathrm{d}k'
\end{align}

A further elaboration of \eref{discrete} requires the specification of the scattering rates $W_{kk'}$ and $W_{k'k}$. In this case, for simplicity, we will account just for one collision process of one electron with one phonon. In particular, we will consider that, because of the collision, the initial wave vector of the electron $k_0$ changes to a final value $k_F$. By using the Fermi Golden rule in \eref{scattering_rate}, such electron-phonon interaction can be associated to the terms:

\begin{subequations}
\begin{align}
\label{W_{kk}1}
W_{kk'}&=\alpha \delta(k'-k_0)\lim_{\sigma \to 0}e^{-\frac{(k-k_F)^2}{\sigma^2}} \\
W_{k'k}&=\alpha \delta(k'-k_F)\lim_{\sigma \to 0}e^{-\frac{(k-k_0)^2}{\sigma^2}}
\label{W_{kk}2}
\end{align}
\end{subequations}
where the parameter $\alpha$ takes into account all (irrelevant for our simple example) details of the specific computation of the Fermi Golden rule. The parameter $\sigma \to 0$ means that rates $W_{kk'}$ and $W_{k'k}$ are localized closely to momentum $k=k_F$ and $k=k_0$, respectively. For numerical reasons, we avoid writing explicitly delta functions in the right hand side of  \eref{W_{kk}1} and \eref{W_{kk}2}. In simple words, $W_{kk'}$ is the transition rate associated to an electron initially in $k'=k_0$ that appears finally at $k=k_F$ (see $2^{nd}$ term in Fig. \ref{scattering_process1}), and $W_{k'k}$ is associated to an electron initially in $k=k_0$ that finally disappears from $k_0$ (see $3^{rd}$ term in Fig. \ref{scattering_process1}). The summary of this scattering process described in \eref{discrete}  (or in the sum in Fig. \ref{scattering_process1}) is just that an electron with initial momentum $\hbar k_0$ gets a final momentum to $\hbar k_F$ because of the interaction with a phonon. This is the quantum version of the classical collision explained in \sref{sec3}.

Substituting the scattering rates written in  \eref{W_{kk}1} and \eref{W_{kk}2} into Eq.(\ref{discrete}) and rearranging it, we obtain:
\begin{align}
\label{discrete2}
F_W(x,k,t_S+\Delta t)&=F_W(x,k,t_S)\nonumber\\
 &+ \frac{\alpha \Delta t}{2 \pi}F_W(x,k_0,t_S)e^{-\frac{(k-k_F)^2}{\sigma^2}}\nonumber\\
 &-\frac{\alpha \Delta t}{2 \pi}F_W(x,k,t_S)e^{-\frac{(k-k_0)^2}{\sigma^2}}
\end{align} 
Since there is a one-to-one correspondence between the Wigner distribution function and the density matrix \cite{Wigner1932,Case}, one can obtain the density matrix by the inverse Wigner-Weyl transform of the Wigner distribution function as:
\begin{equation}
\label{density_matrix}
\hat \rho(x,x',t)=\int_{-\infty}^{\infty} F_W(\frac{x+x'}{2},k,t)e^{ i k (x-x')} \mathrm{d}k
\end{equation}
As a consequence, we can rewrite \eref{discrete2} as,
\begin{align}
\label{density_matrix1}
\hat \rho(x,x',t_S+\Delta t)&=\hat \rho_B(x,x',t_S)   \nonumber\\
& +  \frac{\alpha \Delta t}{2 \pi}F_W(\frac{x+x'}{2},k_0,t_S)e^{ik_F(x-x')} \nonumber\\
 &- \frac{\alpha \Delta t}{2 \pi}F_W(\frac{x+x'}{2},k_0,t_S)e^{ik_0(x-x')}
\end{align}
which can be rewritten in the form of positive and negative summands discussed in \eref{wig2} as:
\begin{eqnarray}
\label{density_total} 
\hat \rho(t)&=&\hat \rho_B(t) + \hat \rho_P(t)-\hat \rho_N(t)\nonumber\\
&=&|\psi_B \rangle \langle \psi_B | + |\psi_P \rangle \langle \psi_P|-|\psi_N \rangle \langle \psi_N|
\end{eqnarray}
where the first term in the right-hand side, $\hat \rho_B$, describes the density matrix before the collision, the second and third terms are the terms generated (by the Boltzmann Collision operator) due to the collisions. It is important to underline that we are selecting $\alpha$ small enough to ensure that the charge density of the density matrix $\hat \rho(x,x',t_S+\Delta t)$ in \eref{density_matrix1} is strictly non-negative everywhere just after the scattering. As commented previously, it would be a nonsense to subtract more probability presence than what we have in one specific location at $t_S$. Even with this important requirement, the problem may appear later when the time evolution of  $\hat \rho_B(x,x',t_S)$ and $- \frac{\alpha \Delta t}{2 \pi}F_W(\frac{x+x'}{2},k_0,t_S)e^{ik_0(x-x')}$ in \eref{density_matrix1} (or $-|\psi_N \rangle \langle \psi_N|$ in \eref{density_total}) becomes different. 

\begin{figure}[h]
\begin{minipage}{18cm}
  \centering
          \subfloat[]{\label{wigner11}
          \includegraphics[width=0.45\textwidth,height=0.27\textwidth]{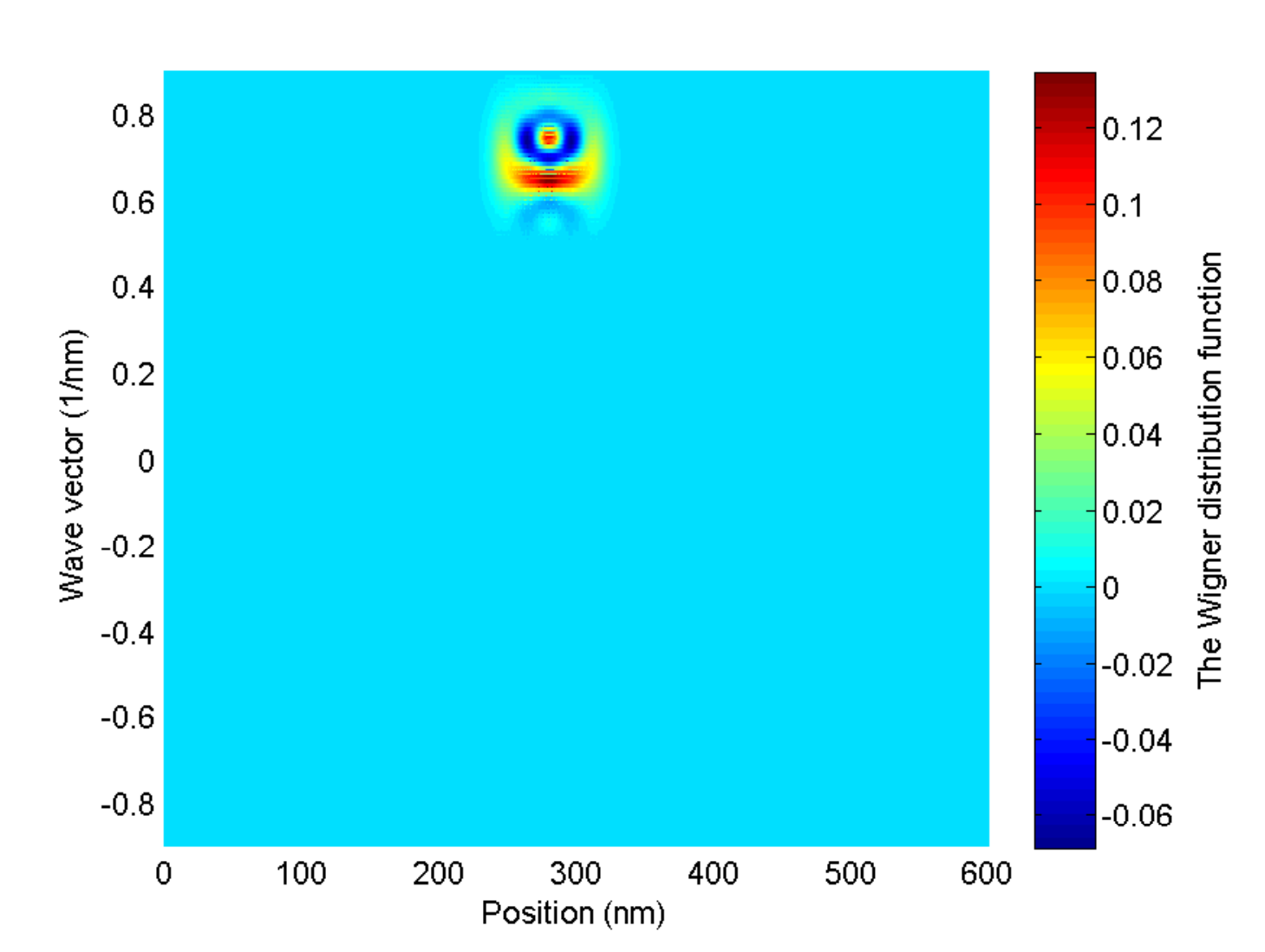}}
           \subfloat[]{\label{charge11}
          \includegraphics[width=0.45\textwidth,height=0.27\textwidth]{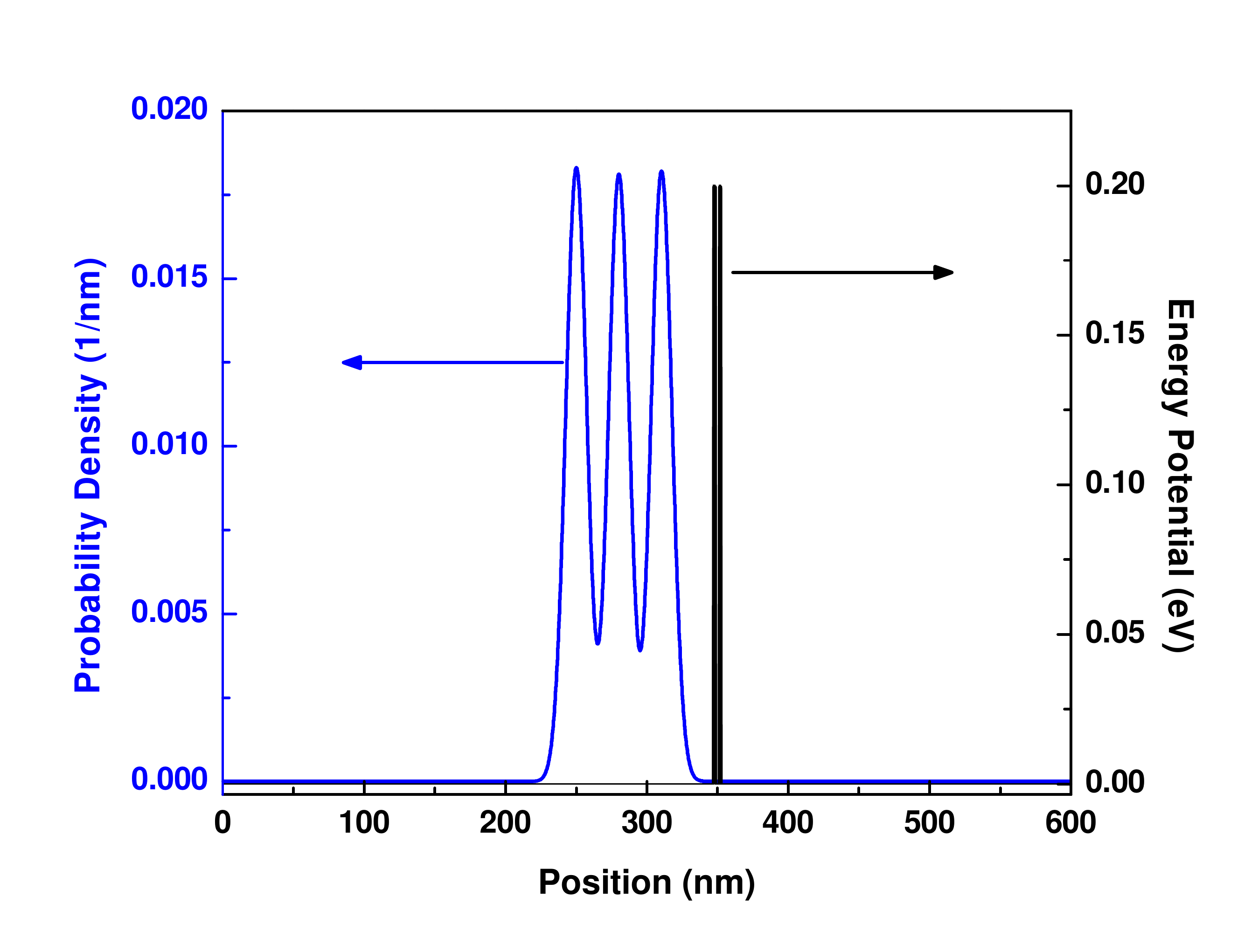}}\quad
          \subfloat[]{\label{wigner11}
          \includegraphics[width=0.45\textwidth,height=0.27\textwidth]{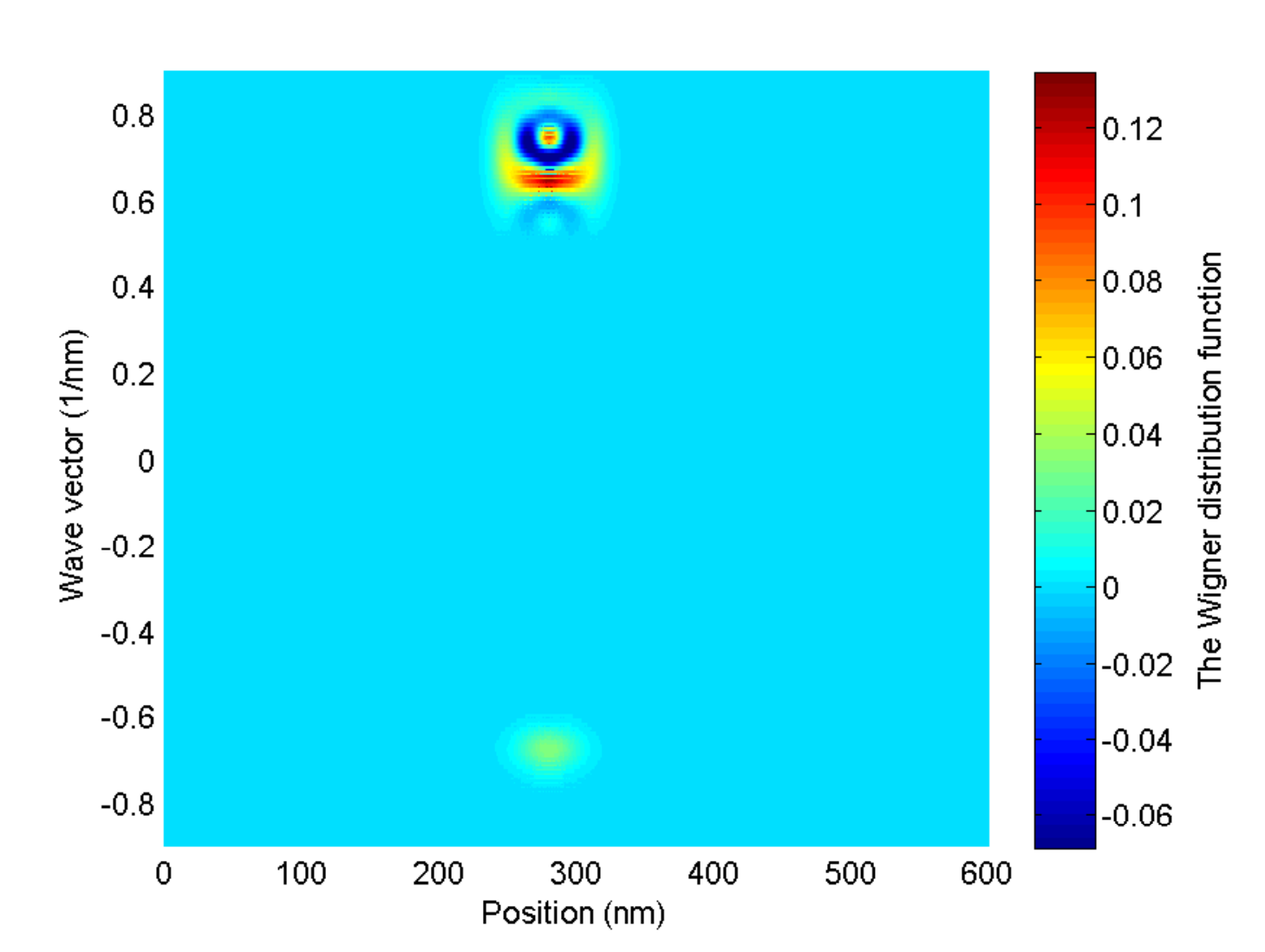}}
           \subfloat[]{\label{charge11}
          \includegraphics[width=0.45\textwidth,height=0.27\textwidth]{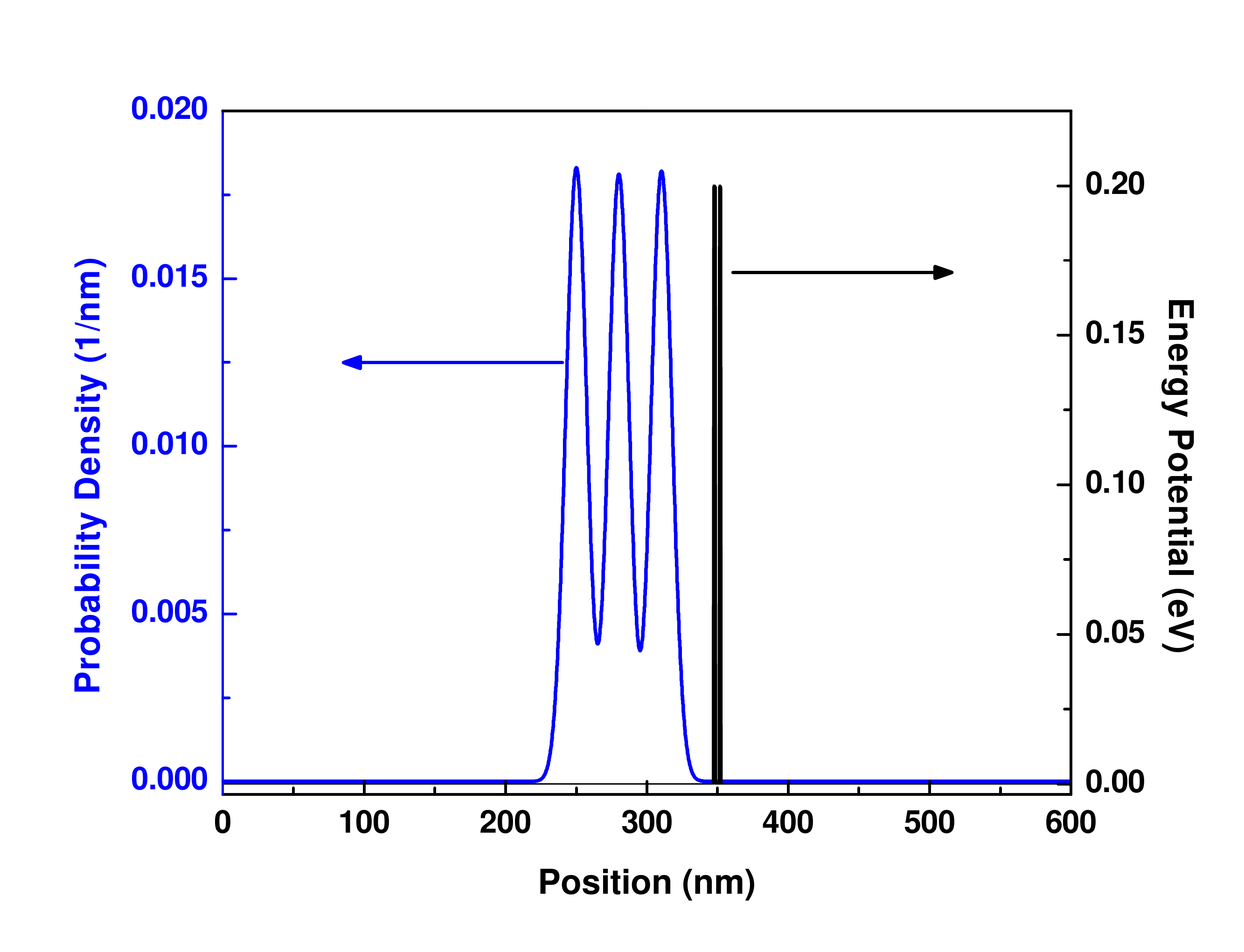}}\quad
          \subfloat[]{\label{wigner11}
          \includegraphics[width=0.45\textwidth,height=0.27\textwidth]{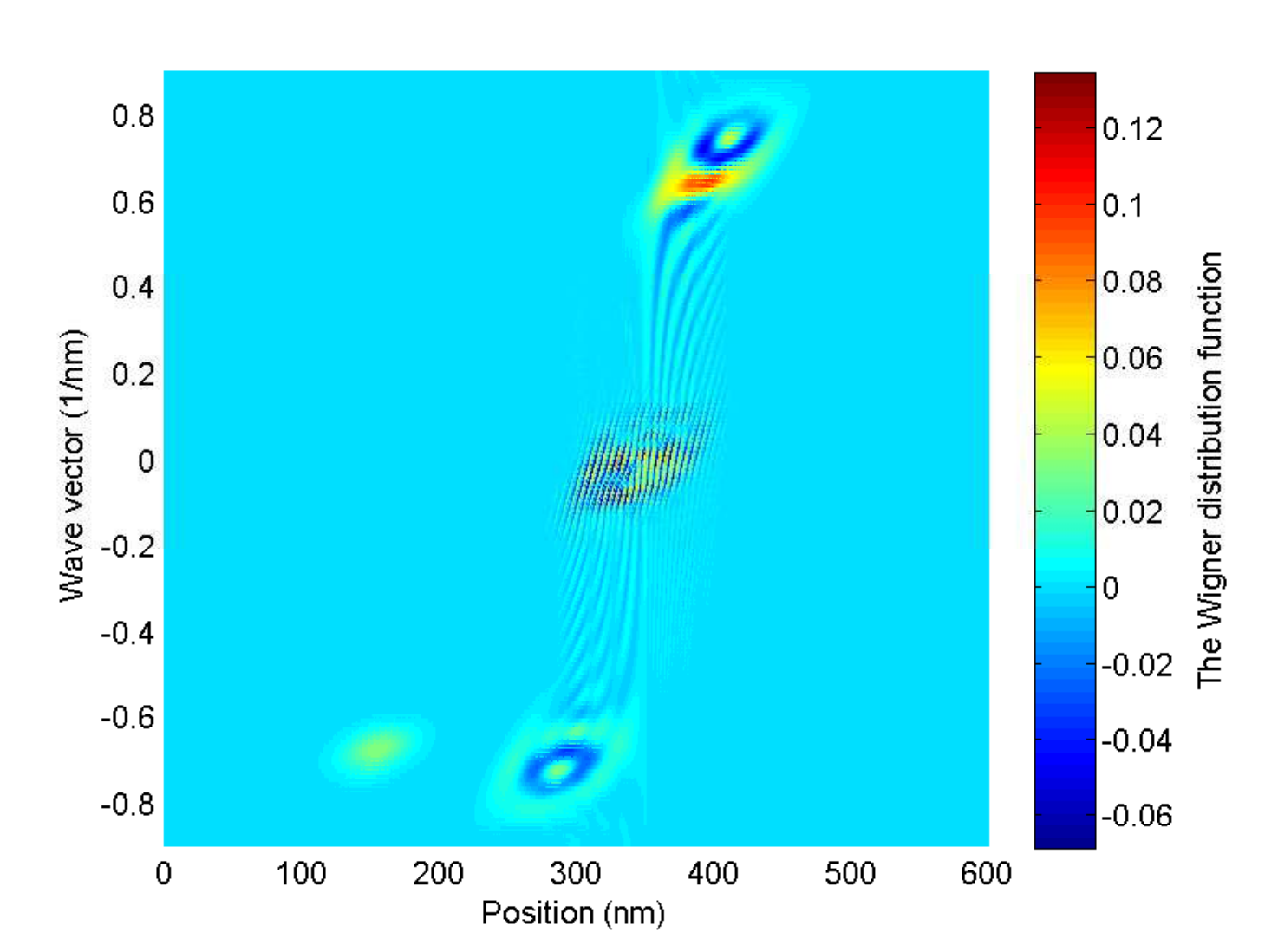}}
           \subfloat[]{\label{charge11}
          \includegraphics[width=0.45\textwidth,height=0.27\textwidth]{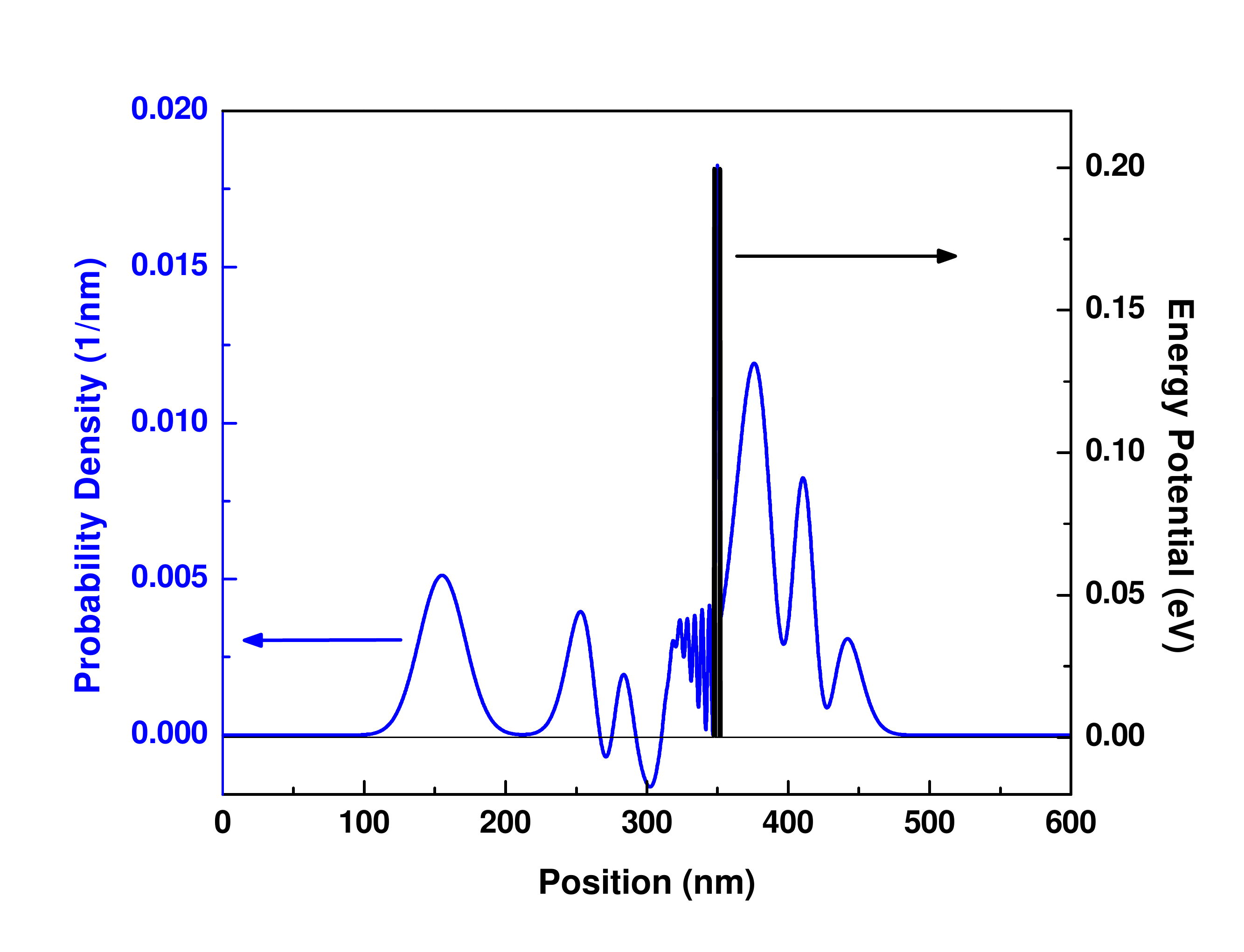}}\quad
          \subfloat[]{\label{wigner11}
          \includegraphics[width=0.45\textwidth,height=0.27\textwidth]{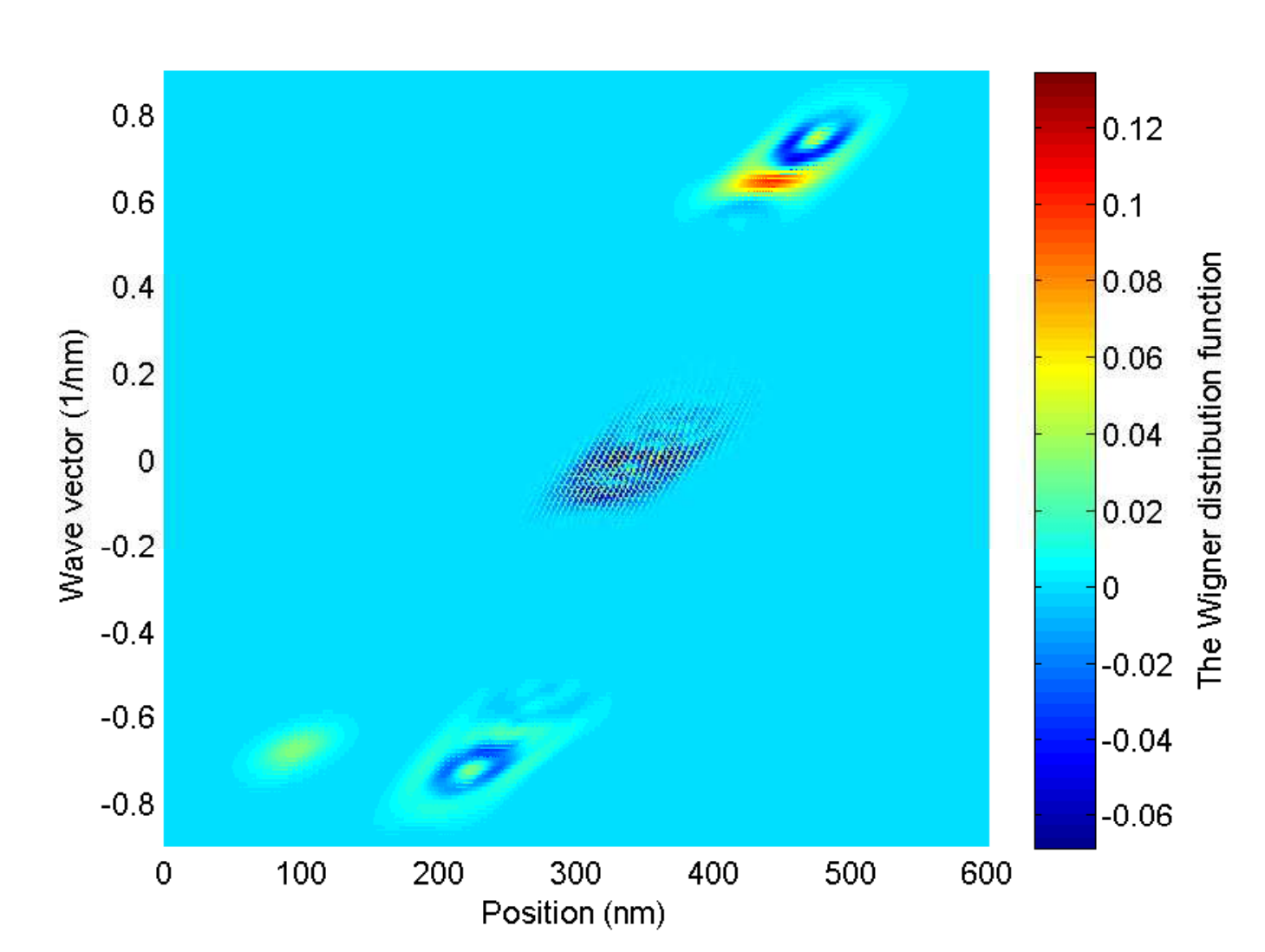}}
           \subfloat[]{\label{charge11}
          \includegraphics[width=0.45\textwidth,height=0.27\textwidth]{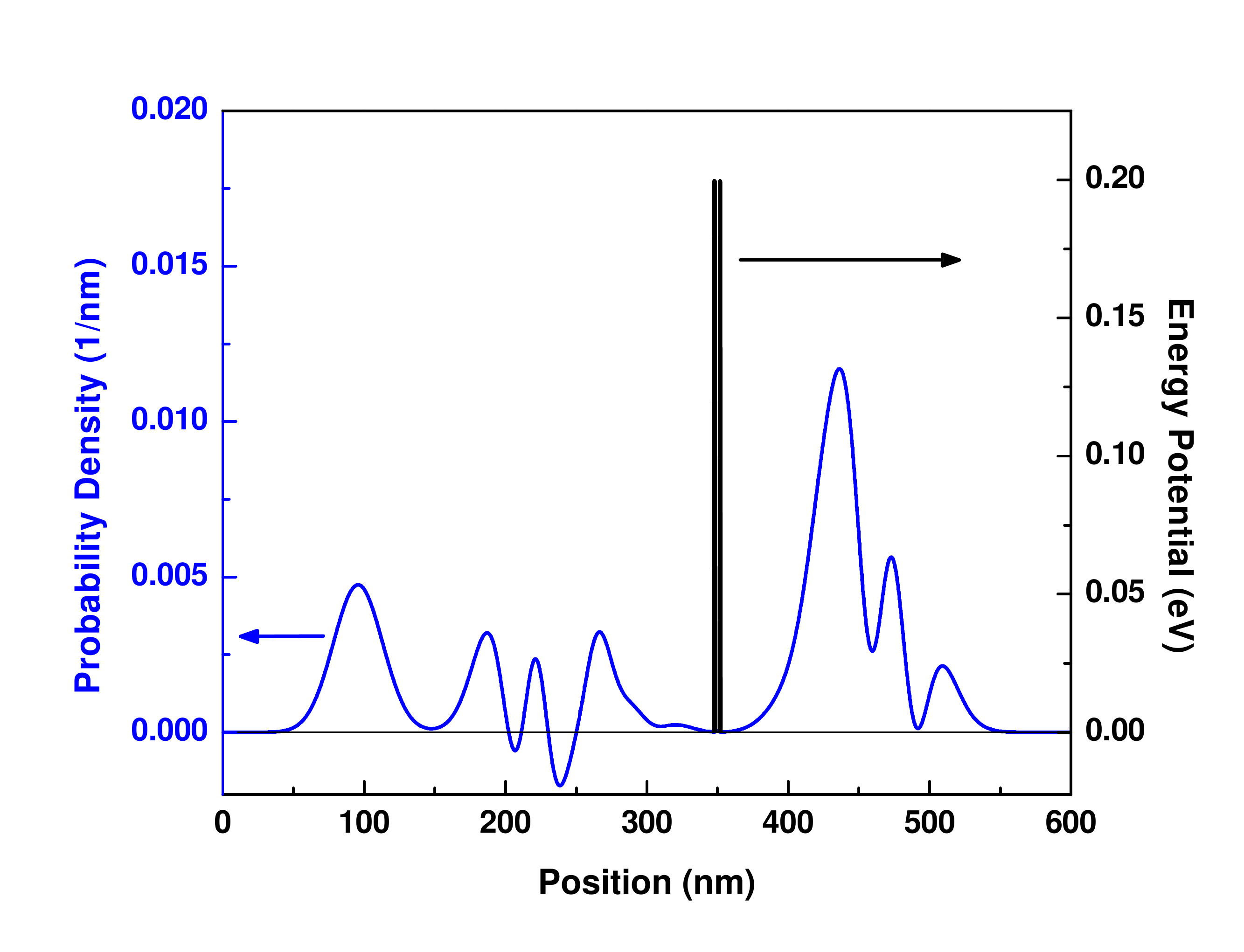}}
     \caption{(Color online) Evolution of Gaussian wave packets coupled with the Hamiltonian eigenstates scattering approach moving towards barriers. (a), (c), (e), (g) are the Wigner distribution function and (b), (d), (f), (h) are the corresponding probability (charge) density at four different times: initial time $t_1=0$ ps, scattering time $t_S=0.006$ ps before touching the barriers, time $t_2=0.315$ ps when wave packets are interacting with the barriers and time $t_3=0.66$ ps when the interaction is completely done. The simulation parameters are: $E=0.09$ eV, $m^*\!=\!0.2\;m_0$, where $m_0$ is the free electron mass, the barrier height is $0.2$ eV, the barrier width is $0.8$ nm and the well depth is $4$ nm.}
  \label{plane_wave}
 \end{minipage}
\end{figure}

We compute the charge probability distribution from Eq.(\ref{marg2}) at four different times corresponding to the initial time $t_0=0$ ps, just after the scattering time $t_S=0.006$ ps, at $t_2=0.315$ ps when the wave packets $\psi_B$ and $\psi_N$  are interacting with the barriers, and at time $t_3=0.66$ ps when the interaction is nearly finished and the initial wave packets $\psi_B$ and $\psi_N$ are clearly split into transmitted and reflected components. The information corresponding to these four times is plotted in Fig. \ref{plane_wave}.

In order to enlarge the typical interference effects,  at the initial time $t_0$ we consider the following initial state $\langle x|\psi_B\rangle=C \langle x|\psi_1+\psi_2+\psi_3 \rangle$ with $C$ a normalization constant. Each wave function  $\psi_j(x,t_0)$ at the initial time $t_0$ is a Gaussian wave packet $\psi_j(x,t_0)=(\frac{2}{\pi a_0^2})^{\frac{1}{4}}e^{i k_0(x-x_{0j})} exp \left( -\frac{(x-x_{0j})^2}{a_0^2} \right)$ but with different initial central positions $x_{0j}$. In particular, the left wave packet $\psi_1 $ has $x_{01}=250$ nm, the middle wave packet $\psi_2 $ has $x_{02}=280$ nm and the right wave packet is $\psi_3$ has $x_{03}=310$ nm. The initial spatial variance of the three wave packets is $a_0=15$ nm, its central wave vector $k_0=0.69\; \textrm{nm}^{-1}$ and the effective mass $m^\ast =0.2 \;m$ with $m$ being the free electron mass. The center of the barriers is at $x=350$ nm. Both barriers have a $0.8$ nm width, the energy height is $0.2$ eV, and they are separated by 4 nm. 

After the scattering process, the evolution of the whole density matrix in \eref{density_total} implies the new term $\psi_P$ with new momentum $\hbar k_F$  (associated to the density matrix is  $\hat \rho_P$ in \eref{density_total}) and the new terms $\psi_N$ with the old momentum $ \hbar k_0$ (associated to the density matrix $\hat \rho_N$ in \eref{density_total}). The two additional wave packets $\psi_N$ and $\psi_P$ are Gaussian wave packets with the same very large dispersion $a_{0S}=2a_0\bigg(\sqrt{1+\frac{4 \hbar^2 t_S^2}{m_\ast^2 a_0^4}}\bigg)$ (to mimic a plane wave) and the same central position $x_{0S}=x_0+\frac{\hbar k_0}{m_\ast}t_S$. The wave vectors for $\psi_N$ and $\psi_P$ are $k_0$ and $-k_0$ (here we assume $k_F=-k_0$), respectively.

The results of the charge (or probability presence) densities in the right hand side of the Fig. \ref{plane_wave} are just \eref{marg2}, that is the integral of the Wigner distribution function in the left-hand side of the figure over all momenta (for a fixed position). The negative values of the Wigner distribution function in the figures is not at all problematic as far as the marginal integral in \eref{marg2} satisfies $Q(x,t) \geq 0$ \cite{Enrique}.  Before and just after the collision, there is not unphysical evolution of the charge. Just after the collision, the new state $|\psi_P\rangle$  gives only positive charge density and the negative contribution of the new state $|\psi_N\rangle$ is, obviously, smaller than the positive one provides by $|\psi_B\rangle$ at each location. Figs.  \ref{plane_wave}(e) and (f) show the same information at the new time $t_2=0.315$ ps when the wave packets $\langle x|\psi_B\rangle$ and $\langle x|\psi_N\rangle$ have evolved with time and interacting with the barriers. At position $x=300$ nm, negative charge (presence probability) appears. This result is unphysical in the same sense that a wave function with a negative modulus will be unphysical, i.e. inconsistent with the probabilistic (Born law) interpretation of quantum mechanics.  After the interaction is completed, at time $t_2=0.66$ ps, the spurious phenomenon becomes even worse. In this particular simple example, there are more positions (for example $x=492$ nm) with negative value probabilities. 

We have also performed simulations (not shown in the present paper) where the scattering process is strictly performed according to \eref{density_matrix1} using $- \frac{\alpha \Delta t}{2 \pi}F_W(\frac{x+x'}{2},k_0,t_S)e^{ik_0(x-x')}$ instead of $|\psi_N \rangle \langle \psi_N|$. The results are quite similar to the ones show in Fig. \ref{plane_wave}, but because of the own positive/negative oscillation of the $F_W(\frac{x+x'}{2},k_0,t)$, the charge probability results are even worse than the ones plotted here. The amplitude where the negative probability occurs is larger and such negative value appears in more positions, which is affected by the factor $\frac{\alpha \Delta t}{2\pi}F_W(x,k_0,t_S)$ presenting the scattering strength. It is important to point out that the unphysical spurious behaviors become even worse with longer time evolutions (related to the device active region). In conclusion, the presence of negative charge is not because of the exact shape of the $\rho_N$ and $\rho_P$, either pure states or mixed states, but because the time-evolution of the states  $\rho_N$ is different from $\rho_B$ because, at some times $t$,  their positive and negative contribution cannot be compensated (even if they were compensated at $t_0$). 

\clearpage
\clearpage


\section{Numerical example of the solution}
\label{sec4}

As indicated in \sref{sec3} the solution to avoid these unphysical results of \fref{plane_wave}, while still using the simple ideas of the Boltzmann collision operator, is having an exact knowledge of the states involved in the description of the density matrix. By construction, this way of working will not provide any negative charge density. 

\subsection{Boltzmann collision operator for general states}

Let us assume that we have a perfect knowledge of all states, $|\psi_i \rangle$ with $i=1,...,N$, that are needed to build our density matrix of the open system in \eref{wig1}. We define first $F^i_W(x,k,t)$ as the Wigner-Weyl transform with respect to the element of the density matrices $ p_i|\psi_i \rangle \langle\psi_i|$ in \eref{wig1}. Then, because of the linearity of the Wigner distribution function with respect to the density matrix,  before the collision, we can write the whole Wigner distribution function as follows:
\begin{eqnarray}
F_W(x,k,t)= \sum_{i=1}^N F^i_W(x,k,t)
 \label{wig4}
\end{eqnarray}
Inspired in the classical application of the Boltzmann collision operator, we define a collision operator in  \eref{Boltzmann_operator} that provides transitions between different states  as:

\begin{align}
\label{Boltzmann_operator2}
&\hat{C}_W\left[ F_W(x, k, t)\right] = \nonumber\\
&\qquad\quad \frac{1}{2 \pi}\sum_{i=1}^{N^*}\sum_{j=1}^{N^*}\{Z_{ij}F_W^j(x, k, t)-Z_{ji}F_W^i(x, k, t)\}
\end{align}  
where the terms $Z_{ij}$ provides the scattering rate (for the \emph{general} states used in each case) from the $j-$ state $|\psi_j \rangle$ to the $i-$ state $|\psi_i \rangle$. The sums in \eref{Boltzmann_operator2} are carried out over the $N^*$ possible existent terms (which are in principle infinite, but we can limit to a reasonable number of possible states in a practical application). We do not use $W_{kk'}$ in \eref{Boltzmann_operator2} because, in principle, they are computed only for Hamiltonian eigenstates, while we define $Z_{ij}$ using our general states $|\psi_j \rangle$\footnote{As a reasonable approximation, if the general wave packet has (a more or less) well defined momentum (for example, the mean momentum of the wave packet) the terms $W_{kk'}$ can be numerically used instead of $Z_{ij}$ in \eref{Boltzmann_operator2} by using some relations between $i$ (and $j$) and $k$ (and $k'$).}.

\subsection{Electron in a double barrier with a collision event}

\begin{figure}[t!!!!!!]
\centering
\includegraphics[scale=0.5]{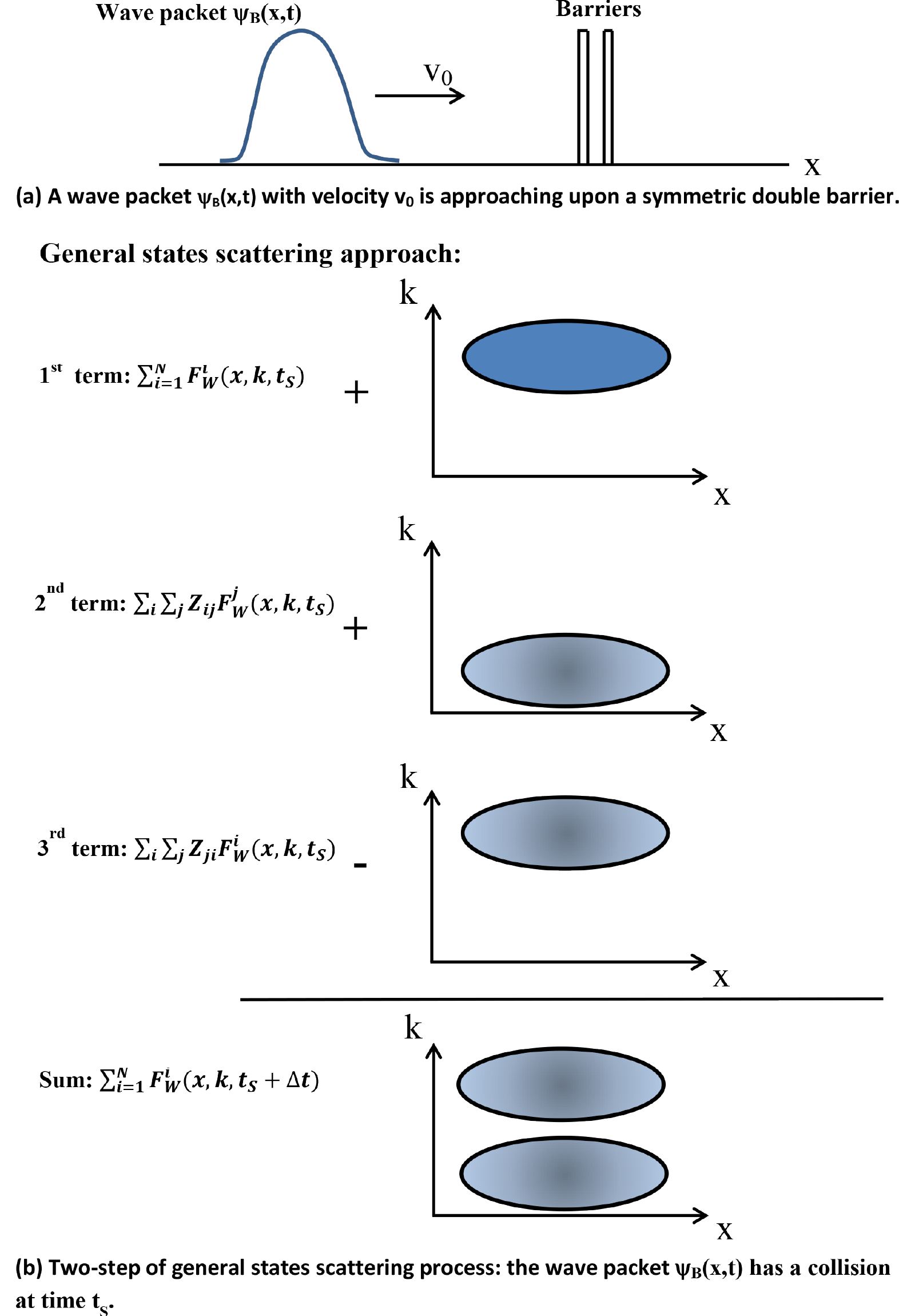}
\caption{Schematic representation of the two-step general states scattering process. (a) Simulation of a wave packet impinging on double barriers, the collision is performed at time $t_S$ before the wave packet touches the barriers. (b) A simple physical picture of the two-step general states scattering process.}
\label{scattering_process2}
\end{figure}

We discuss here the same numerical example presented in \sref{sec2}, but here with our new general collision operator in \eref{Boltzmann_operator2}. We use the same initial density matrix  $\hat{\rho}_B(t_0)= |\psi_B \rangle \langle\psi_B|$ in \eref{wig1}. We consider that there are two electrons with such state, $M_1=M=2$. Then, when the scattering take place, one of the two electrons with initial state $|\psi_B \rangle$ changes its state, while the other remains unaffected.   The new density matrix in \eref{wig2} is $\hat{\rho}(t_S)=\hat{\rho}_B-(1/2)|\psi_B \rangle \langle\psi_B|+ (1/2)|\psi_{F} \rangle \langle\psi_{F}|$. The new density matrix after scattering can be greatly simplified to $\hat{\rho}(t_S)=(1/2)|\psi_B \rangle \langle\psi_B|+(1/2)|\psi_{F} \rangle \langle\psi_{F}|$ because $|\psi_N \rangle \equiv \frac{1}{\sqrt{2}}|\psi_B \rangle$. This collision process is explained in \fref{scattering_process2}.

\begin{figure}[h]
\begin{minipage}{18cm}
  \centering
          \subfloat[]{\label{wigner11}
          \includegraphics[width=0.45\textwidth,height=0.27\textwidth]{wigner0.pdf}}
           \subfloat[]{\label{charge11}
          \includegraphics[width=0.45\textwidth,height=0.27\textwidth]{charge_t10.pdf}}\quad
          \subfloat[]{\label{wigner11}
          \includegraphics[width=0.45\textwidth,height=0.27\textwidth]{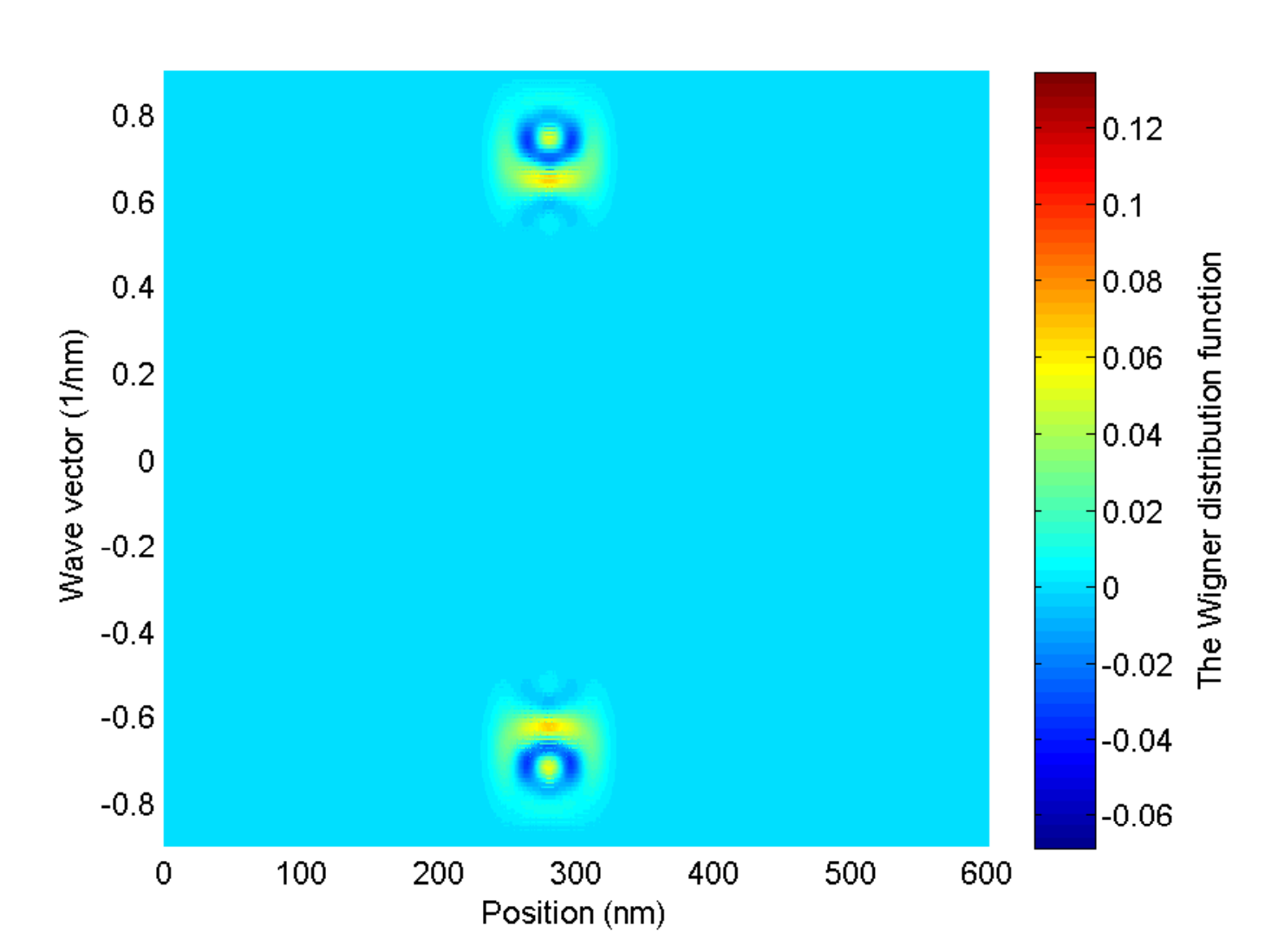}}
           \subfloat[]{\label{charge11}
          \includegraphics[width=0.45\textwidth,height=0.27\textwidth]{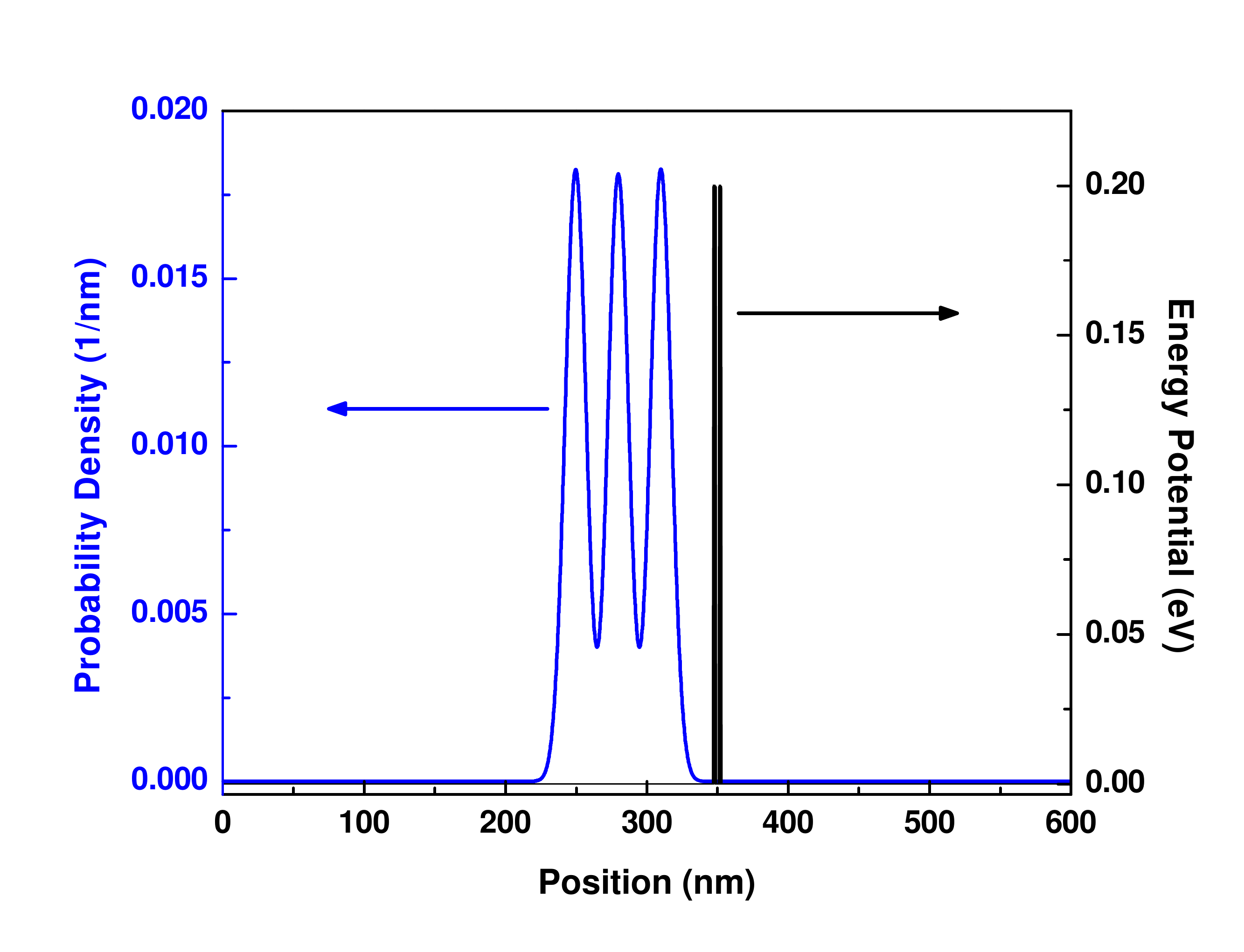}}\quad
          \subfloat[]{\label{wigner11}
          \includegraphics[width=0.45\textwidth,height=0.27\textwidth]{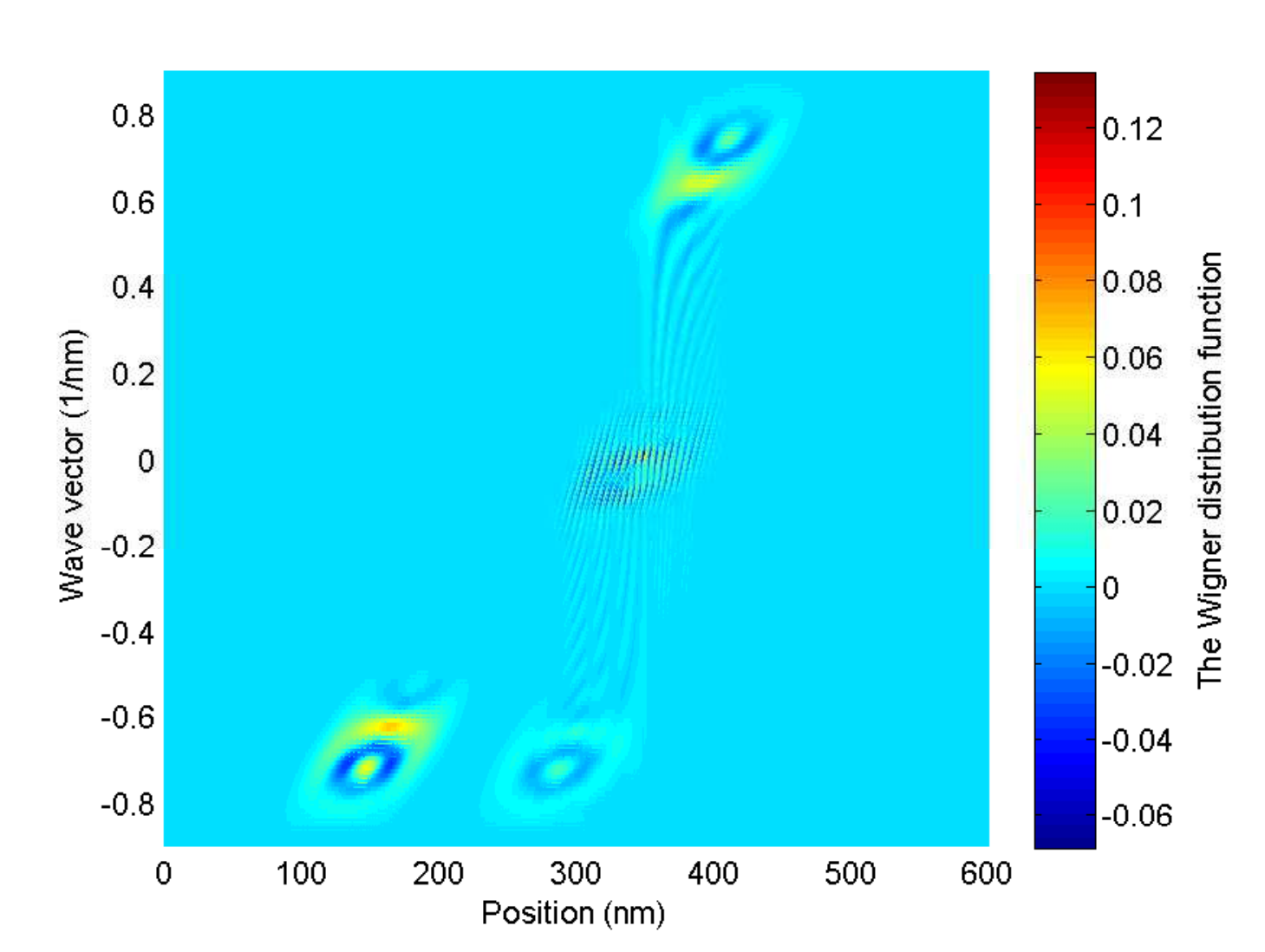}}
           \subfloat[]{\label{charge11}
          \includegraphics[width=0.45\textwidth,height=0.27\textwidth]{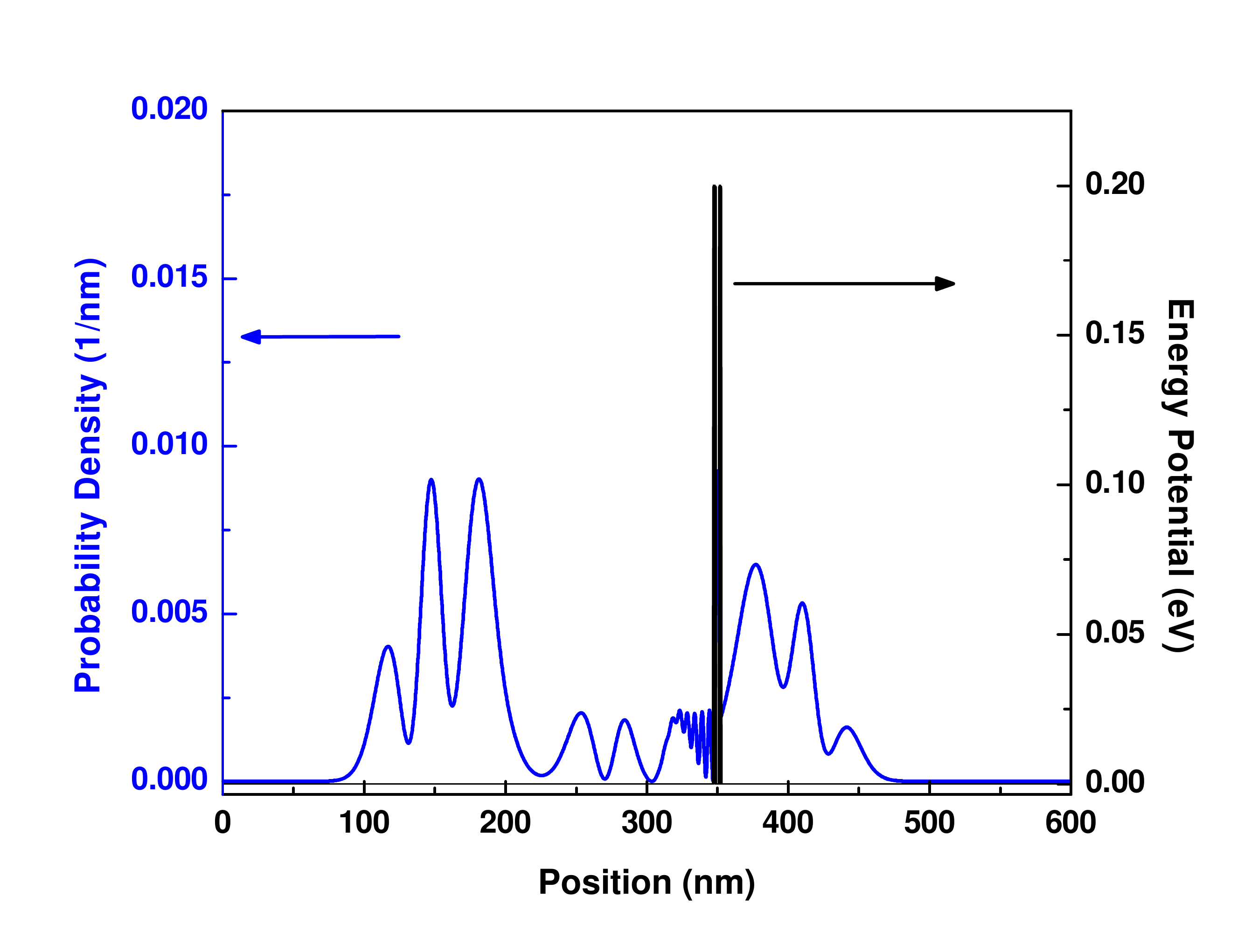}}\quad
          \subfloat[]{\label{wigner11}
          \includegraphics[width=0.45\textwidth,height=0.27\textwidth]{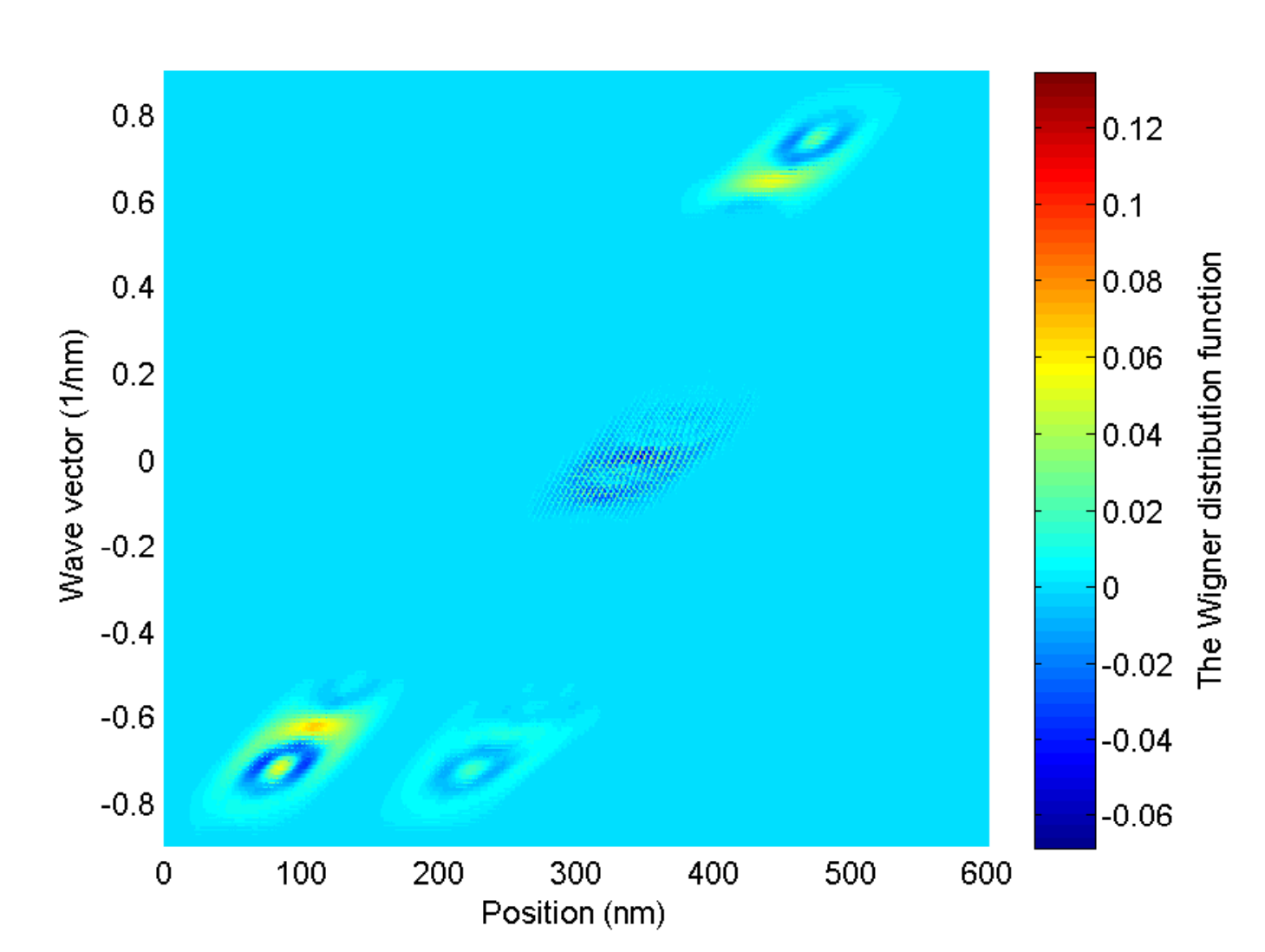}}
           \subfloat[]{\label{charge11}
          \includegraphics[width=0.45\textwidth,height=0.27\textwidth]{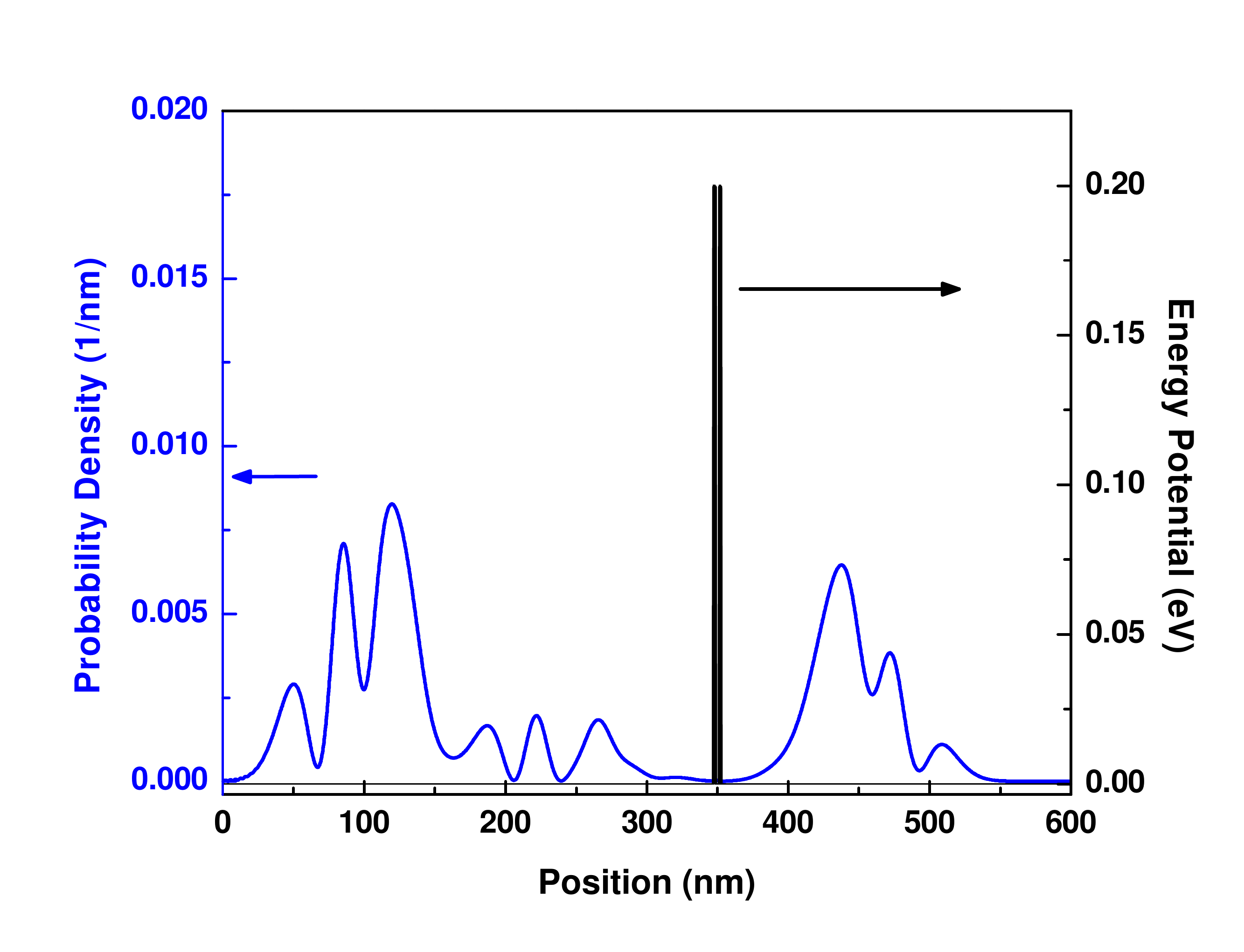}}
     \caption{(Color online) Evolution of Gaussian wave packets coupled with general states scattering approach moving towards barriers. (a), (c), (e), (g) are Wigner distribution and (b), (d), (f), (h) are corresponding probability density of state at for different times, which is identical to the time in \fref{plane_wave}: $t_1=0$ ps, $t_S=0.006$ ps, $t_2=0.315$ ps and $t_3=0.66$ ps. The simulation parameters are: $E=0.09$ eV, $m^*\!=\!0.2\;m_0$, where $m_0$ is the free electron mass, the barrier height is $0.2$ eV, the barrier width is $0.8$ nm and the well depth is $4$ nm.}
  \label{wp}
 \end{minipage}
\end{figure}

In our numerical example, we use the same initial wave packet $|\psi_B \rangle$ discussed in \fref{plane_wave}. At the initial time $t_1=0$ ps, the information of the Wigner function and the charge probability distribution plotted in \fref{wp} are identical to that in \fref{plane_wave}. Then, at time $t_S=0.006$ ps, the new collision operator in \eref{Boltzmann_operator2} acts on the Wigner distribution function. After the scattering, the system state is $\hat \rho=|\psi_B \rangle \langle \psi_B | + |\psi_P \rangle \langle \psi_P|-|\psi_N \rangle \langle \psi_N|=\frac{1}{2}|\psi_B \rangle \langle \psi_B|+\frac{1}{2}|\psi_F \rangle \langle \psi_F|$. As a consequence, the negative values disappear. Therefore, the unphysical results are removed. These conclusions are perfectly corroborated by \fref{wp}. 
 
We remark also that this new algorithm for collision explained here is relevant for time-dependent modeling of quantum transport. In addition, since it requires a perfect knowledge of the states that built the density matrix, its practical implementation fits perfectly well with the BITLLES simulator developed with conditional wave functions \cite{OriolsBook,Oriols2007,Oriols2010,Oriols2013}. Then, the mentioned  algorithm for dissipative quantum transport can be straightforwardly implemented for quantum transport by directly including the interaction between electrons and phonons in the kinetic part of the Hamiltonian that describes the wave function of each electron. For a preliminary description of this method, see Refs. \cite{IWCE2015,Oriols2016}. 

\clearpage
\clearpage

\section{Conclusions}
\label{sec5}

\begin{table}
\begin{tabular}{l*{4}{c}r}
\hline
              &  \multicolumn   {3}{ c }{Norm} \\
\hline
               & Positive & Negative & Total\\ 
               & probability& probability& probability\\
\hline
H.E. scattering           & 1.025 & -0.025 & 1  \\
G.S. scattering           &1 & 0 & 1  \\
\hline
\end{tabular}
\caption{Norm of the system state (positive, negative and total probability density) at $t_3=0.66$ ps with coupling to the Hamiltonian eigenstates (H.E.) and general states (G.S.) scattering}
\label{Norma}
\end{table}

In this work we have discussed the combination of the Boltzmann collision operator and the Fermi Golden rule to simulate dissipative processes in quantum electron devices. In general, the Fermi golden rule provides the transition rates $W_{kk'}$ and $W_{k'k}$ for (unperturbed) Hamiltonian eigenstates. We have shown that this combination can provide negative unphysical values for the charge (probability) in some simple scenarios, as we proved in \sref{sec2} (Fig. \ref{plane_wave}).  

We have also discussed the ultimate reasons of such unphysical results: the different time-evolutions of the states that build the density matrix in the Wigner distribution function and the Hamiltonian eigenstates generated by the Boltzmann collision operator after a collision. This problem will never occur if the transition between states during the collision process is done according to the set of states that built the density matrix. In Table \ref{Norma}, the numerical results show negative values of the charge density when the Boltzmann collision operator is applied with Hamiltonian eigenstates.

In principle, the unphysical results can be avoided by using more elaborated forms of the collision operator \cite{Jonasson}. However, then, the intuitiveness and the computational simplicity offered by the Boltzmann collision operator would be missed. We mentioned in \sref{sec4} that by construction, an approach based on the generalization of the Boltzmann collision operator as written in \eref{Boltzmann_operator2} will always avoid such unphysical feature. An algorithm based on these ideas will still retain the intuitiveness and the computational simplicity of the Boltzmann collision operator. However, it will require the knowledge of the states that conform the Wigner distribution function. This detailed knowledge of the states that build the density matrix is trivially accessible in a formulation of quantum transport in terms of conditional wave functions. See, for instance, the BITLLES simulator \cite{BITLLES}. Then, the collision process can be applied directly into the Hamiltonian of the equation of motion that determines the evolution of the conditional wave function (just adding an additional term in the kinetic part of the Hamiltonian). For  more details, see \cite{IWCE2015} and \cite{Oriols2016}.

\begin{acknowledgements}
This work has been partially supported by the \lq\lq{}Fondo Europeo de DesarrolloRegional (FEDER) and Ministerio de Economia y Competitividad through the Spanish Project TEC2015-67462-C2-1-R and by European Union Seventh Framework Program under the Grant agreement no: 604391 of the Flagship initiative  \lq\lq{}Graphene-Based Revolutions in ICT and Beyond\rq\rq{}. Z. Zhan acknowledges financial support from the China Scholarship Council (CSC).
\end{acknowledgements}

%

\end{document}